\documentclass[footinbib,twocolumn,amsmath,amstex,amssymb,mathfonts,superscriptaddress,prl]{revtex4}
\usepackage{graphicx}
\usepackage{color}
\usepackage{bm}

\usepackage{amsmath}
\usepackage{amssymb}
\usepackage{amsthm}
\usepackage{amsfonts}
\usepackage{multirow}
\usepackage{makecell}
\usepackage{float}
\usepackage{comment}
\usepackage{enumitem}
\usepackage{verbatim}
\usepackage{mathtools}
\usepackage{esint}
\usepackage{tipa}
\usepackage{braket}
\usepackage{verbatim}
\usepackage{stackengine}

\newcommand\eqdef{\stackrel{\text{def}}{=}}
\newcommand\CF{_{\text{\tiny{CF}}}}

\begin{document}

\title{Spin-statistics relation and the Abelian braiding phase for anyons in fractional quantum Hall effect}
\author{Ha Quang Trung, Yuzhu Wang and Bo Yang$^*$} 
\affiliation{Division of Physics and Applied Physics, Nanyang Technological University, Singapore 637371.}
\email{yang.bo@ntu.edu.sg}

\date{\today}
\begin{abstract}
Quasihole excitations in fractional quantum Hall (FQH) systems exhibit fractional statistics and fractional spin, but how the spin-statistics relation emerges from many-body physics remains poorly understood. Here we prove a spin-statistics relation using only FQH wave functions, on both the sphere and disk geometry. In particular, the proof on the disk generalizes to all quasiholes in realistic systems, which have a finite size and could be deformed into arbitrary shapes. Different components of the quasihole spins are linked to different conformal Hilbert spaces (CHS), which are nullspaces of model Hamiltonians that host the respective FQH ground states and quasihole states. Understanding how the intrinsic spin of the quasiholes is linked to different CHS is crucial for the generalized spin-statistics relation that takes into account the effect of metric deformation. In terms of the experimental relevance, this enables us to study the effect of deformation and disorder that introduces an additional source of Berry curvature, an aspect of anyon braiding that has been largely neglected in previous literature.
\end{abstract}

\maketitle 

\textit{Introduction--}
Anyons are theoretically proposed particles in two dimensions whose adiabatic exchange leads to a phase between 0 and $\pi$, interpolating the statistics of bosons and fermions\cite{leinaas1977theory,wilczek1982magnetic, preskill1999lecture}. The topological nature of fractional statistics promises applications in robust quantum computing\cite{preskill1998fault,preskill1999lecture,sarma2005topologically,nayak2008non}. One promising platform to realize anyons is the fractional quantum Hall (FQH) systems, where anyons are elementary excitations\cite{arovas1984fractional,leinaas2002spin,yoshioka2002quantum,stern2008anyons,feldman2021fractional}. The fractional statistics first demonstrated in the quasiholes of the Laughlin state\cite{arovas1984fractional,wilczek1982magnetic} motivated the search for an anyonic ``topological spin" that satisfies the standard spin-statistics theorem\cite{goldhaber1976connection,wilczek1982quantum,halperin1984statistics,thouless1985remarks,li1992spin,leinaas2002spin,feldman2021fractional}. However, a complete understanding of this relation remained elusive\cite{leinaas2002spin,feldman2021fractional}. Moreover, the scarcity and difficulty of accurate measurements in experiments of these fractional braiding phases\cite{camino20073,an2011braiding,willett2019interference, nakamura2020direct, bartolomei2020fractional,glidic2022cross} raise the question of how robust this statistical phase is in realistic experimental conditions.

\begin{figure}
\begin{center}
\includegraphics[width=\linewidth]{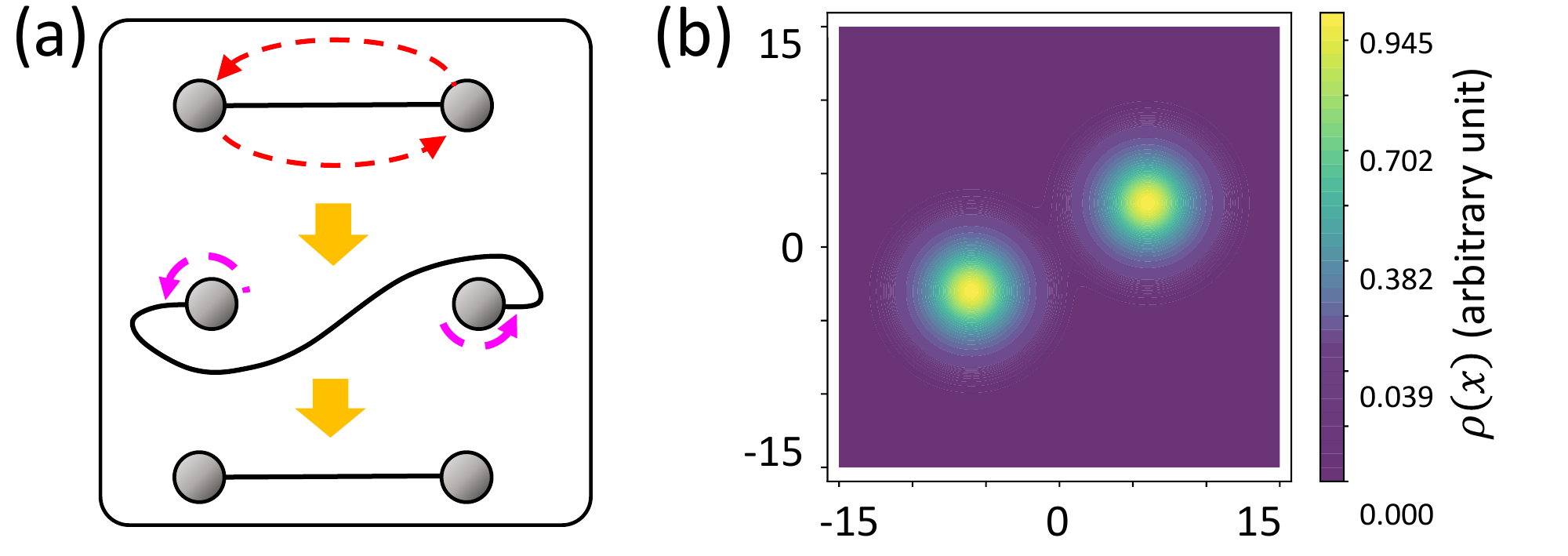}
\caption{(a) The ``attached string" argument: exchanging the positions of two particles must be followed by self-rotation. (b) Two particles (electrons or anyons) in a single LL slightly deforms each other, allowing a well-defined ``string attachment".}
\label{string argument}
\end{center}
\end{figure}

The proper understanding of the anyonic spin-statistics relations is important both theoretically and experimentally. Formally, such relationship for elementary particles requires the full machinery of the relativistic quantum field theory (r-QFT)\cite{pauli1940connection,sakurai2014modern,mund2009spin,baez1995topological}. In its most simplified form, Lorentz invariance (for bosons) and the additional requirement of energy being bounded from below (for fermions) are needed\cite{pauli1940connection}. On the other hand, nonrelativistic explanations employ the intuitive picture that exchanging two particles involves particle self-rotation\cite{duck1998toward,preskill1999lecture}. This argument relies on a ``string attachment"\cite{duck1998toward} as shown in Fig. \ref{string argument}a. While the existence of such a string cannot be physically justified for point particles, quasiparticles in condensed matter systems are not point particles\cite{johri2014quasiholes} (see Fig. \ref{string argument}b). Furthermore, since Lorentz invariance is mostly irrelevant in such systems, we can in principle expect a nonrelativistic justification for the spin-statistics relation\cite{baez1995topological,preskill1999lecture,leinaas2002spin, hansson1992dimensional}.

In the context of FQH anyons, the standard spin-statistics relation, $\gamma_{exc}=2\pi s$, fails (here $\gamma_{exc}$ is the phase gained after exchanging two particles and $s$ is their intrinsic spin. In this paper we use the term ``exchange" to refer to the process of exchanging two anyons on the 2D plane, while ``braiding" to refer to the process of winding one anyon around a second stationary anyon; thus for a given Abelian anyon species, the braiding phase is twice the exchange phase.). For an Abelian state at filling factor $\nu$, the intrinsic spin of a cluster of $k$ quasiholes (also referred to as ``$k$-stacked"), obtained by parallel transport on the sphere\cite{goldhaber1976connection, halperin1984statistics,li1992spin, read2008quasiparticle,li1992spin, einarsson1995fractional,leinaas2002spin}, or from electron density on the disk\cite{comparin2022measurable,umucalilar2018time}, is shown to take the form
\begin{equation}
\label{total am}
s_k = -\frac{\nu k^2}2+\frac{k}{2}+n\nu k.
\end{equation}
where $n$ is the Landau level (LL) index. It clearly contradicts the well-known exchange statistics of $\nu\pi$\cite{arovas1984fractional}(e.g. for $k=1$, $\nu=1$, $\gamma_{exc}=\pi$ but $2\pi s_k=2\pi n$).
 
Experimental studies of the anyon statistics are further complicated by the coupling of the quasihole charge to the magnetic field. The Berry phase from the exchange of two anyons in QH systems is $\gamma_{\text{Berry}}=\gamma_{\text{B-field}}+\gamma_{\text{stats}}$ where $\gamma_{\text{B-field}}$ is the Aharonov-Bohm (AB) phase from the background magnetic field, and the $\gamma_{\text{stats}}$ encodes the exchange statistics of the anyons. Probing the latter quantities experimentally requires observing a discrete change in the total Berry curvature\cite{nakamura2020direct, carrega2021anyons, willett2019interference} or measuring the scattering amplitude from anyon colliders\cite{bartolomei2020fractional,glidic2022cross}. In both cases the quasiholes propagate along the edge, which has been shown to be robust for certain simple cases\cite{feldman2022robustness}. However, a microscopic understanding of the effect of disorder on anyonic statistics is lacking, especially for quasiholes that reside completely in the bulk. A very important aspect of anyons is that they are \emph{not point particles}, but occupy a finite area with a shape that can be easily deformed. While previous studies have drawn attention to the correction to the braiding phase due to the overlapping tails of the quasiholes\cite{kjonsberg1997anyon,kjonsberg1999charge,simon2008effect,rosenow2016current,sondhi1992long}, even in the large separation limit where the quasiholes are well-separated, deforming the shapes of the quasiholes themselves can lead to further correction to the Berry phase. This has important implications to the experimental braiding of anyons and the potential realisation of robust universal topological quantum computers.

In this Letter, we propose an intuitive and rigorous understanding for the relationship between the intrinsic spin and statistics for all types of Abelian anyons. These include not only the special cases of the fermions and bosons, but also anyons from $k$-stacked quasiholes that occupy a finite area, as well as those with \emph{an arbitrary shape} (e.g. deformed by external potential or disorder). A modification to the standard spin-statistics relation, which generalizes to the braiding of a $k$-stack around a $k'$-stack, has been proposed as\cite{einarsson1995fractional,comparin2022measurable,li1992spin}:
\begin{equation}
\label{general spin-statistics}
\theta_{k,k'}=2\pi(s_k+s_{k'}-s_{k+k'})
\end{equation}
Here we prove it analytically by treating the adiabatic exchange as a perturbation to some rotationally invariant Hamiltonian. Further more, in the case where $k=k'$ (identical particles) it is natural to define a physically relevant intrinsic spin, of which the topological spin\cite{einarsson1995fractional,read2008quasiparticle} is the special case, so that the standard spin-statistics relation is recovered. We also discuss the effect of disorder on the Berry phase measurement on a quantum Hall (QH) droplet. While our discussion was limited to quasiholes in Abelian phases, we can easily generalise to Abelian quasielectrons\cite{yang2021elementary,wang2021geometric} and Abelian braiding of non-Abelions\cite{seesup,macaluso2019fusion}, and in principle the intrinsic spin is well-defined for such cases.

\textit{$k$-stacked quasiholes on the sphere--}
We first look at a QH fluid realised on the spherical geometry with a magnetic monopole at the center\cite{haldane1983fractional,greiter2011landau}, generalising the original arguments in Ref.\cite{li1992spin}. Inserting a $k$-stacked quasihole into the ground state of a generic quantum Hall phase with $N_e$ electrons results in the total number of fluxes $N_{\phi}=\nu^{-1}N_e-s_c-s_f+k$, where $s_c, s_f$ are the cyclotron and guiding center topological shifts\cite{wen1992classification,park2014guiding}. We can then take the difference of the adiabatic rotation phase about the z-axis between the two states, a $k$-stacked quasihole at the north pole ($\gamma_1$) and at the south pole ($\gamma_2$)\cite{seesup}, from the difference between their angular momentum:
\begin{eqnarray}
&&\Delta\gamma_{21}=\gamma_2-\gamma_1=2\pi k\nu N_{\phi}+4\pi\left(\gamma_{s_1}+\gamma_{s_2}+\gamma_{s_3}\right)\quad\label{phase12a}\\
&&\gamma_{s_1}=\frac{\nu k}{2}s_c,\quad\gamma_{s_2}=\frac{\nu k}{2}s_f\quad\gamma_{s_3}=-\frac{\nu}{2}k^2\label{phase12b}
\end{eqnarray}
The first term on the RHS captures the coupling of the $k-$stacked anyon to the total magnetic flux. The additional terms are understood as the parallel transport on the sphere inducing a self-rotation of the quasihole due to the presence of the Gaussian curvature\cite{wen1992classification,einarsson1995fractional,read2008quasiparticle}. For the special case of the Laughlin state, $s_f=\nu^{-1}-1$ and the three terms in Eq.(\ref{phase12b}) sums up to Eq.(\ref{total am}). 

On the disk this quantity corresponds to the Berry phase of a $k$-stack moving in an infinitely large circle\cite{fano1986configuration}. The north pole is mapped to the center of the disk while the south pole is mapped to infinity. A $2\pi$ rotation of the quasiholes about the $z$-axis on the sphere corresponds to their adiabatic dragging in a circular loop on the disk. It is important to note that only $\Delta\gamma_{21}$ is measurable as the AB phase since the curvature is zero everywhere on the disk. This is contrary to the naive view that, since a quasihole moving in a circular loop on the disk is equivalent to rotating the entire system, the associated Berry phase should be given by only $\gamma_2$\cite{leinaas2002spin, li1992spin}. Subtracting from $\gamma_2$ the angular momentum of the neutral ground state\cite{leinaas2002spin} gives the angular momentum of the quasihole\cite{comparin2022measurable}, but still does not give the right phase. Instead, the measurable quantity is the \emph{excess} angular momentum from that of the quasihole placed at the rotation center. 

Microscopically this can be understood by considering a realistic system where the quasihole is trapped and rotated by some potential profile. Consider a local trapping potential $\hat H(\theta) = \hat H_0 + \hat H_1(\theta)$ where $\hat H_1$ is a $\theta-$dependent perturbation to the rotationally invariant trapping potential $\hat H_0$. The ground state $|\psi(\theta)\rangle$ can be written as
\begin{align}
|\psi(\theta)\rangle &= \lambda_0|\psi_0\rangle+\lambda_1|\psi_1(\theta)\rangle,\quad \langle\psi_0|\psi_1(\theta)\rangle=0\label{perturb1}
\end{align}
where $|\psi_0\rangle$ is the ground state of $\hat H_0$. Tuning $\theta$ rotates the ground state, giving the Berry connection\cite{seesup} $A_\theta = -\left(\langle\psi\left(\theta\right)|L_z|\psi\left(\theta\right)\rangle-\langle\psi_0|L_z|\psi_0\rangle\right)$. This depends only on the \emph{excess} angular momentum. Intuitively, we cannot physically rotate a perfectly symmetric object on a flat surface. Only a deformed quasihole can be rotated, and the Berry phase comes from the \emph{change} in angular momentum caused by deformation.

\begin{figure}
\begin{center}
\includegraphics[width=\linewidth]{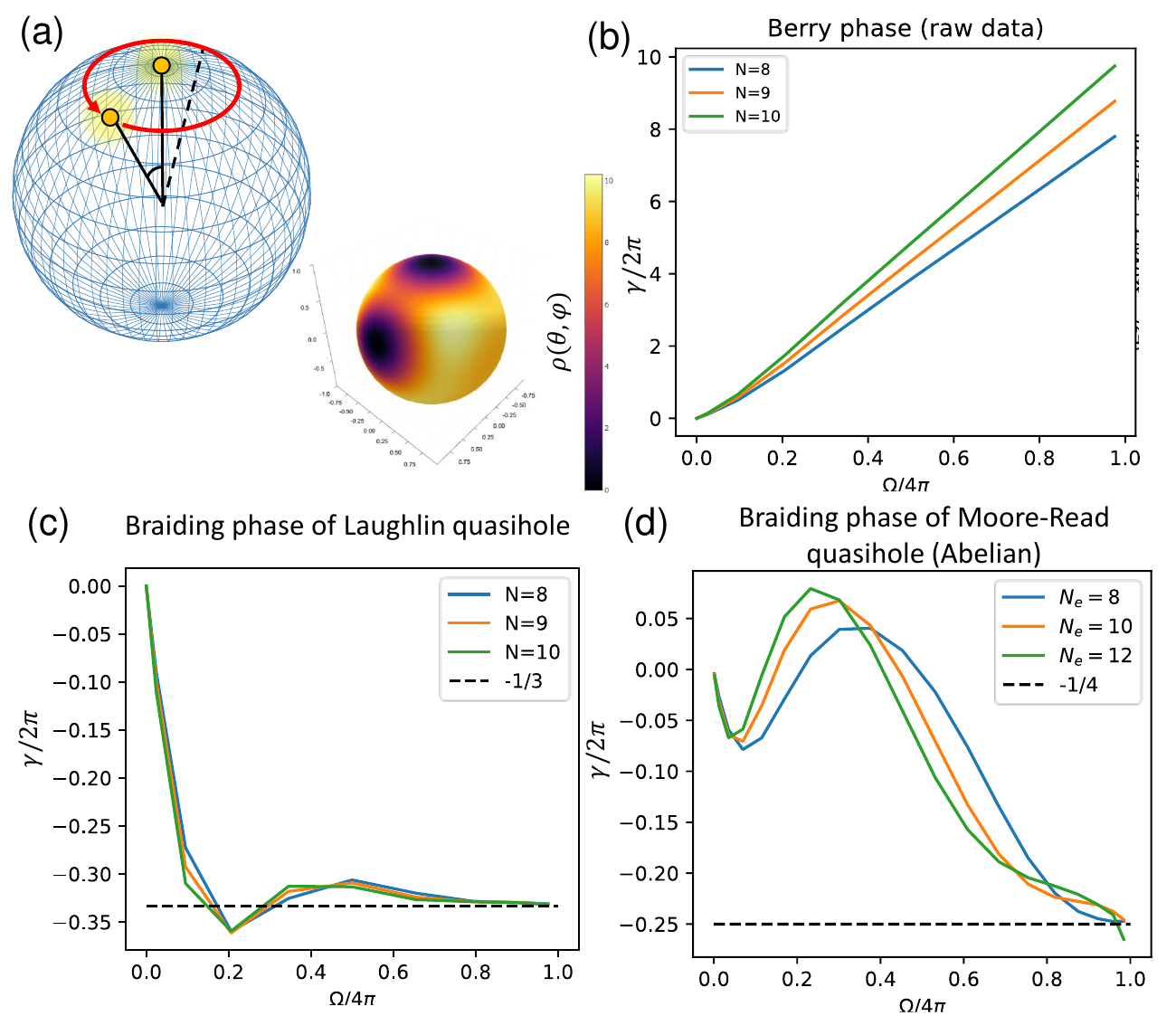}
\caption{(a) Schematic diagram of the braiding process on the sphere. The inset shows a sample heatmap of the electron density on the sphere calculated for a two-quasihole Laughlin state. (b) Total Berry phase for braiding two Laughlin quasiholes shows an $\Omega$-dependent part (linear as $\Omega\to4\pi$) and an $\Omega$-independent part. The latter can be extracted as the braiding phase. (c) Braiding phase of two Laughlin quasiholes calculated for different system sizes. (d) Braiding phase of two Abelian Moore-Read quasiholes (two half-fluxes) calculated for different system sizes.}
\label{fig:sphere braiding}
\end{center}
\end{figure}

\textit{The derivation of the spin-statistics theorem--} 
Let us now start by inserting $\left(k+k'\right)$ fluxes to an FQH ground state at the north pole, creating a $\left(k+k'\right)$-stacked anyon there. Next we pull $k$ of the fluxes to the south pole, leaving behind $k'$ fluxes at the north pole. The $2\pi$ rotation gives the phases $\gamma_3=-\left(k+k'\right)N_e\pi, \gamma_4=\left(k-k'\right)N_e\pi$ respectively. The phase difference between these two scenarios is\cite{seesup}:
\begin{eqnarray}
\Delta\gamma_{43}&=&\gamma_4-\gamma_3=\Delta\gamma_{21}-2\pi\nu kk'\label{phase34}\\
&=&\Delta\gamma_{21}-2\pi(s_k^{topo}+s_{k'}^{topo}-s_{k+k'}^{topo})\label{braiding2}
\end{eqnarray}
where $\Delta\gamma_{21}$ is the single $k$-stacked anyon contribution from Eq.(\ref{phase12a}). The extra term is the braiding phase re-expressed in Eq.(\ref{braiding2}). Here we define a topological spin $s_k$ compatible with the clustering of anyons\cite{read2008quasiparticle, feldman2021fractional}:
\begin{equation}
\label{topological spin}
s_k^{topo}=-\frac{\nu k^2}2
\end{equation}
This topological spin originates from an \emph{intrinsic spin} that we will justify later. While Eq.(\ref{braiding2}) agrees with Eq.(\ref{general spin-statistics}), our calculation reveals the microscopic origin of each term. For rotationally invariant quasiholes, this formula gives a braiding phase of $2\pi\nu kk'$ for an adiabatic braiding of a $k$-stacked anyon around a $k'$-stacked anyon.

The analytical derivation above is consistant with numerical calculations, which also allow us to go beyond the Laughlin and study other FQH states (see Fig.\ref{fig:sphere braiding}a-c). In Fig.\ref{fig:sphere braiding}d we show some results for the braiding of two Moore-Read quasiholes (charge $e/4$) \cite{an2011braiding,macaluso2019fusion,willett2019interference}. Adding one additional magnetic flux to the Moore-Read ground state gives a quasihole of charge $e/2$, which can be split into two half-fluxes of charge $e/4$ each without any punishment by $V_3^{3bdy}$ (model Hamiltonian for the Moore-Read state). The two-quasihole Hilbert space is isomorphic to the two-quasihole Laughlin Hilbert space, and in principle, one would expect an Abelian braiding. (The non-Abelian property of the Moore-Read state only manifests when there are four or more quasiholes.) For the Moore-Read state, it is difficult to reproduce the analytical calculation as above, since one cannot have a single isolated quasihole. However, the two-quasihole states can be constructed numerically by finding the ground state of two local potentials on the sphere\cite{seesup}. The braiding phase can be extracted as the solid angle-independent part of the total Berry phase when rotating two-quasihole states about the $z$-axis. Our numerical results show the intrinsic spin of the Moore-Read state is still a well-defined quantity and the spin-statistics relation still holds when the quasihole manifold is Abelian. The implication of the intrinsic spin to non-Abelian braiding remains a subject for future study.

The generalised spin-statistics relation can also be understood as a special case of Eq.(\ref{perturb1}), where we deform a rotationally invariant $\hat H_0$, of which the $\left(k+k'\right)$-stacked anyon is the ground state. This deformation pulls a $k$-stacked anyons far away from the center of rotation, and again we rotate the entire system with $\hat H_1$ parametrized by $\theta$, and measuring $\Delta\gamma_{43}$ as the excess angular momentum from the perturbation\cite{seesup}. Given that anyons are not point particles, the ``string attachment" shown in Fig.(\ref{string argument}) is physical even when $\hat H_1$ consists of two well-separated, perfectly circular confining potentials, since their deformed tail, while exponentially suppressed by the separation, allows us to ``attach" the string and track the self-rotation during the exchange. In this picture, the relationship between spin (adiabatic self-rotation) and statistics (adiabatic exchange) can be rigorously established without r-QFT, for any types of Abelian anyons in the QH systems.

\begin{figure}
\begin{center}
\includegraphics[width=\linewidth]{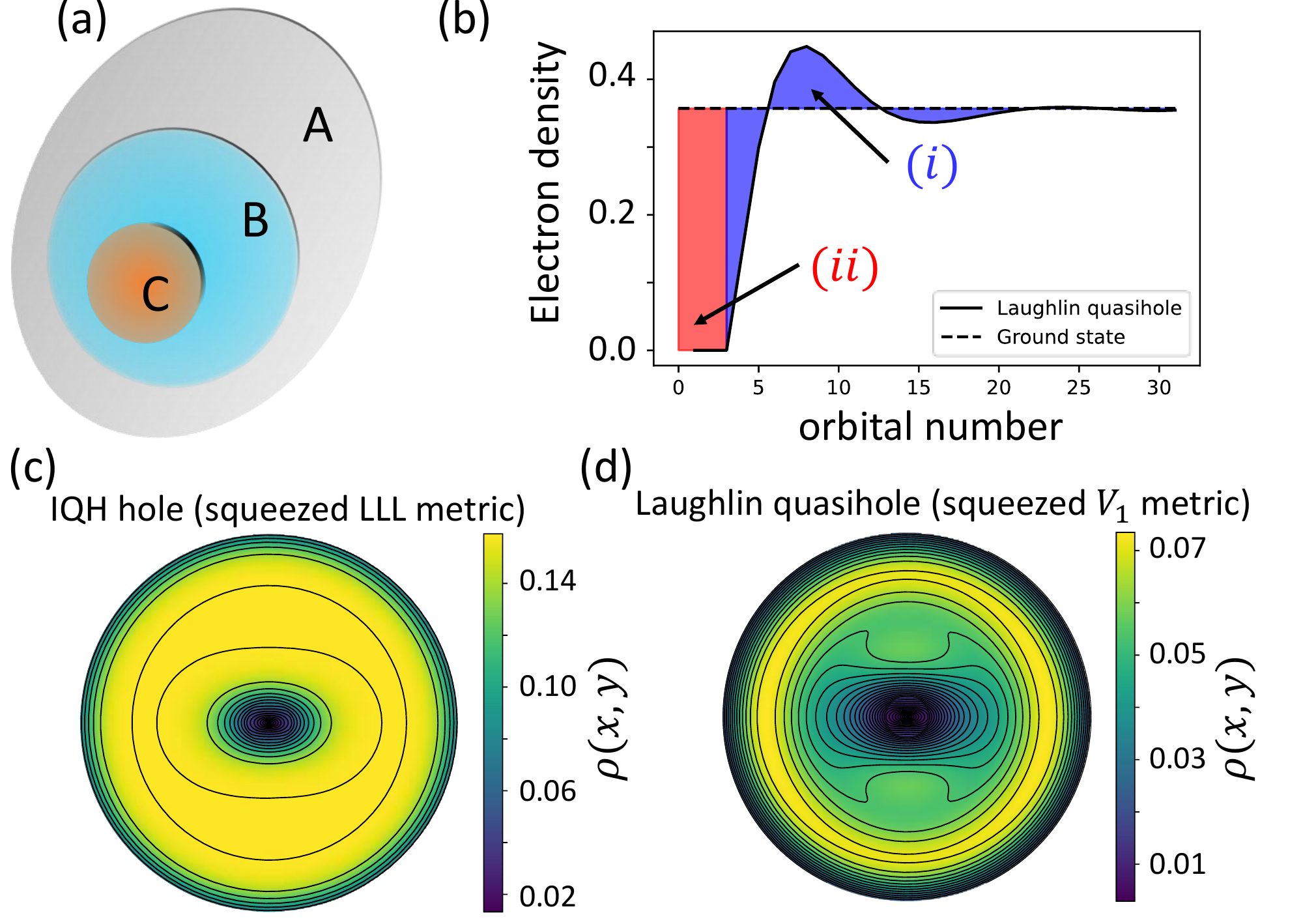}
\caption{(a) CHS hierarchy\cite{wang2021geometric}: $A$ - full many-body Hilbert space,  $B$ - a single LL, $C$ - sub-CHS. (b) Comparing the electron densities of a Laughlin quasihole (solid line) and the neutral ground state (dashed line) shows the origin of $\gamma_{s_2}$ (i) and $\gamma_{s_3}$ (ii). Electron density of (c) the IQH hole squeezed by the guiding center metric and (d) the Laughlin quasihole squeezed by the $\hat V_1^{2bdy}$ nullspace metric\cite{seesup}.}
\label{CHS am}
\end{center}
\end{figure}

\textit{Conformal Hilbert space angular momentum--}
To justify that $s^{topo}_k$ fully captures the Berry phase of the self-rotation of the rotationally invariant $k-$stacked quasihole, we note in Eq.(\ref{phase12b}) the intrinsic angular momentum can be separated into three parts: $L_{\text{cy}}=\gamma_{s_1}$ is the cyclotron angular momentum associated with different LLs; $\gamma_{s_2}$ comes from the FQH topological shift, related to the dipole moment at the edge of the QH fluid\cite{park2014guiding} and vanishes for the IQHE. Here $L_{\text{LL}}=\gamma_{s_2}+\gamma_{s_3}$ is the total guiding center angular momentum within a single LL, as illustrated in Fig.\ref{CHS am}a.

The separation of $L_z$ into $L_{\text{cy}}$ and $L_{\text{LL}}$ is due to $L_{\text{LL}}$ being well-defined within a sub-Hilbert space (a single LL). For any physical operation within a single LL, only $L_{\text{LL}}$ is physically accessible. A single LL is an example of the conformal Hilbert space (CHS) introduced in Ref.\cite{wang2021geometric}
. Thus $L_{\text{LL}}$ is the angular momentum defined within this conformal Hilbert space, characterised by the guiding center metric, or the ``shape" of the CHS. We can deform the shape of a hole within a single LL with a local potential, thus changing the expectation value of $L_{\text{LL}}$, while keeping $L_{\text{cy}}$ invariant, i.e. in the limit of large magnetic field so there is no LL mixing (see  Fig.\ref{CHS am}c). Thus only the deformed guiding center density can be measured. 

Similarly we can further separate $L_{\text{LL}}$ into $\gamma_{s_2}$ and $\gamma_{s_3}$, where the latter is the angular momentum defined within a sub-CHS, the null space of $\hat V_1^{\text{2bdy}}$ interaction denoted as $\mathcal H_1$\cite{wang2021geometric}. Only $\gamma_{s_3}$ is physically relevant if we braid or deform anyons within $\mathcal H_1$ with a local potential much smaller than the incompressibility gap (i.e. $\hat V_1^{\text{2bdy}}$ gives the dominant energy scale), so only the metric characterising $\gamma_{s_3}$ is relevant, while that of $\gamma_{s_2}$ will be invariant. An example is given in Fig.\ref{CHS am}d, showing a Laughlin quasihole deformed by an elliptical potential well in the nullspace of $\hat V_1^{2bdy}$, in complete analogy to Fig.\ref{CHS am}c. For Abelian FQH phases, $\gamma_{s_3}$ can be explicitly computed from a unitary transformation to the composite fermion (CF) basis\cite{yang2022composite,seesup}. By construction the CFs are particles within $\mathcal H_1$, and the $k$-stacked CF holes give the CHS angular momentum of $\gamma_{s_3}$\cite{yang2022composite}.

We are now ready to explicitly write down the general spin-statistics relation:
\begin{equation}
\label{general SSR}
\gamma_{k_1,\eta_1;k_2,\eta_2}=2\pi\left(\bar S_{k_1,\eta_1}+\bar S_{k_2,\eta_2}-s_{k_1+k_2}^{topo}\right)
\end{equation}
giving the phase obtained by adiabatically braiding a cluster of $k_1$ quasiholes around a cluster of $k_2$ quasiholes. Here $\eta_1$ and $\eta_2$ parametrize the deformation, or the internal structure of the quasiholes; $\bar S_{k_i,\eta_i}$ denotes the intrinsic spin of the cluster $k_i$ with deformation $\eta_i$. For rotationally invariant $k_1$-stack, we get $\bar S_{k_1,\eta_1=0}=s_{k_1}^{topo}$ in Eq.(\ref{topological spin}). However, Eq.(\ref{general SSR}) also includes the effect of the internal quasihole structure, encoded in parameters $\eta_1$ and $\eta_2$, which may present an additional contribution to the measured statistics (see subsequent section).
 
For identical particles ($k_1=k_2=k$ and $\eta_1=\eta_2=\eta$), Eq.(\ref{general SSR}) can be simplified to the more familiar form :
\begin{equation}
\label{general SSR identical}
\gamma_{k,\eta} = -4\pi \tilde S_{k,\eta}
\end{equation}
which reminisces the standard spin-statistics relation in rQFT (the minus sign signifies that the quasihole is a deficiency of electrons). For rotationally invariant $k$-stack, $\tilde S_{k,\eta=0}=s_{k}^{topo}=-\nu k^2/2$. For the more general cases (e.g. deformed quasiholes) the intrinsic spin of the quasihole is modified as $\bar S_{k, \eta}=s_k^{topo}+\Delta s_k$, so the braiding phase becomes $-4\pi \tilde S_{k,\eta}=-4\pi\left(s_k^{topo}-\Delta s_k\right)$. The additional factor $-4\pi\Delta s_k$ comes from the self-rotation of the two quasiholes, which becomes non-trivial if their intrinsic shape deviates from the rotationally-invariant wave-packet.


\textit{The deformation of anyons--}
Since each CHS comes with an intrinsic geometric degree of freedom\cite{wang2021geometric}, the quasihole can be arbitrarily deformed while still remaining inside the nullspace of the model Hamiltonian. In practice, this can be done by implementing a trapping potential \emph{within the nullspace} of a given model Hamiltonian, and the metric is determined by the potential profile\cite{seesup}. We consider here an elliptical well that ``squeezes" the quasiholes into elliptical shapes (see Fig.\ref{CHS am}c-d).

\begin{figure}
\begin{center}
\includegraphics[width=\linewidth]{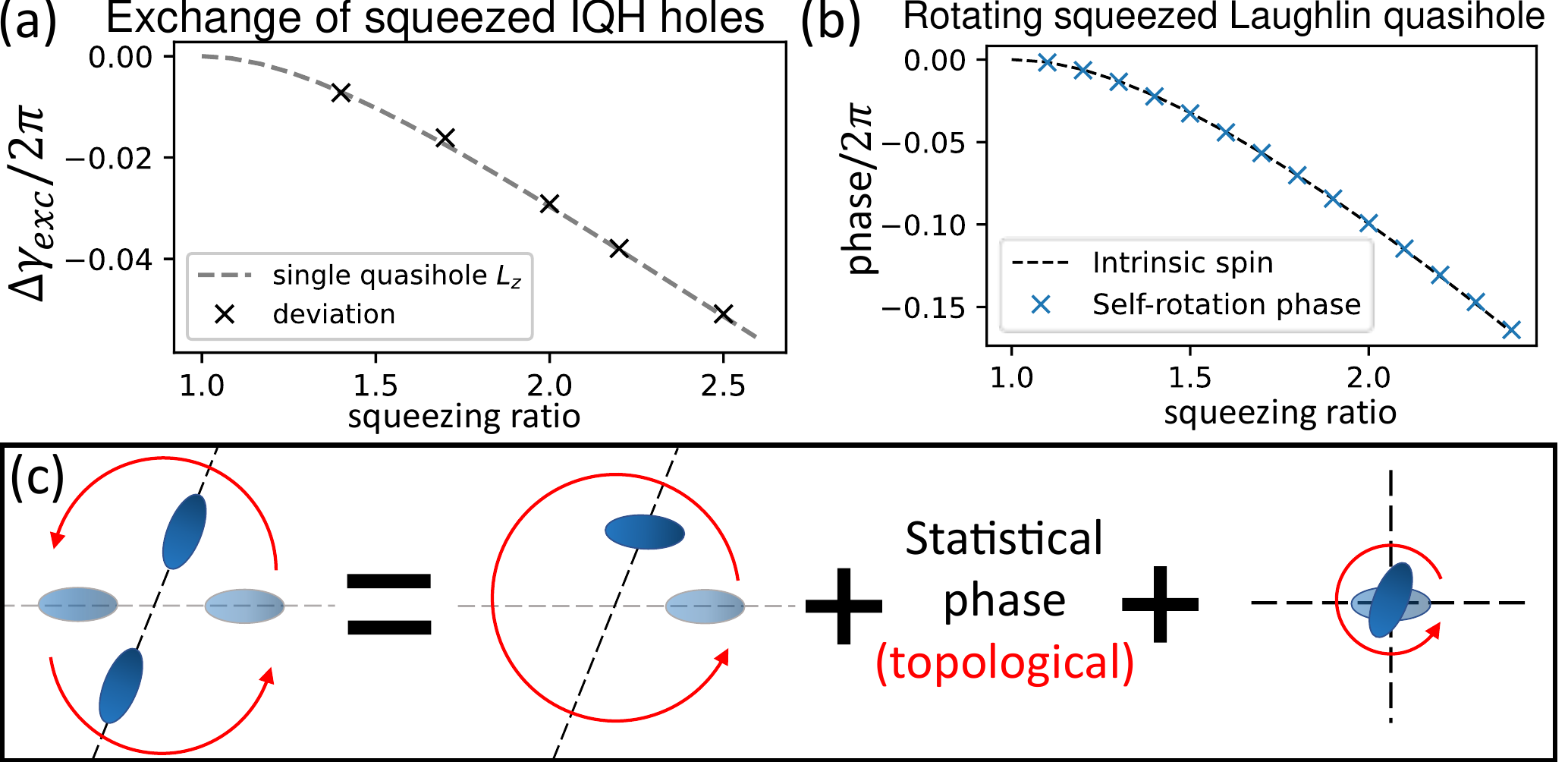}
\caption{(a) Deviation from fermionic statistics at different squeezing ratios (crosses). compared to the deviation of the intrinsic spin from topological spin (dashed line). (b) Self-rotation phase of a single squeezed Laughlin quasihole (crosses) compared to its intrinsic spin (dashed line) at different squeezing ratios. (c) Different contributions to the total Berry phase. The intrinsic spin captures both topological and self-rotation phases.}
\label{ellipse rotate}
\end{center}
\end{figure}

We first illustrate our results with the squeezed holes at filling factor $\nu=1$. Here there exist two possible exchange schemes. When the two holes are exchanged along a circle by pure translation, we observe a fermionic statistics expected of the $\nu=1$ phase\cite{seesup}. However, when each hole self-rotates along the exchange path, there is a deviation as shown in Fig. \ref{ellipse rotate}a, the amount of deviation from fermionic statistics exactly matches the deviation of the intrinsic spin from the topological spin, multiplied by $2\pi$. When self-rotation is involved in the exchange procedure, the exchange phase contains both a topological component that depends only on the topological indices of the FQH phase, and a self-rotation component that depends on the internal shape of the quasiholes (see Fig. \ref{ellipse rotate}c). In order to observe only the topological phase in real experiments, one must ensure that the quasiholes are not rotated by any additional potentials in the system, such as disorder. This is actually difficult to avoid as we will show later.

For FQH states, the physics is analogous once we replace the guiding center angular momentum and the corresponding deformation in the LLL with that of the respective nullspaces. For example with the Laughlin $\nu=1/3$ quasiholes, the intrinsic spin is taken to be the $\hat V_1^{2bdy}$ nullspace angular momentum, exploiting the isometry between the LLL and the $\hat V_1^{2bdy}$ nullspace\cite{wang2021geometric}. This intrinsic spin can be computed from the fermionization process\cite{wang2021geometric,seesup}. Fig. \ref{ellipse rotate}b shows the Berry phase from the self-rotation of a squeezed Laughlin quasihole agrees very well with its intrinsic spin. In particular, Fig.\ref{ellipse rotate}c readily generalises to all Abelian FQH phases.

\begin{figure}
\begin{center}
\includegraphics[width=\linewidth]{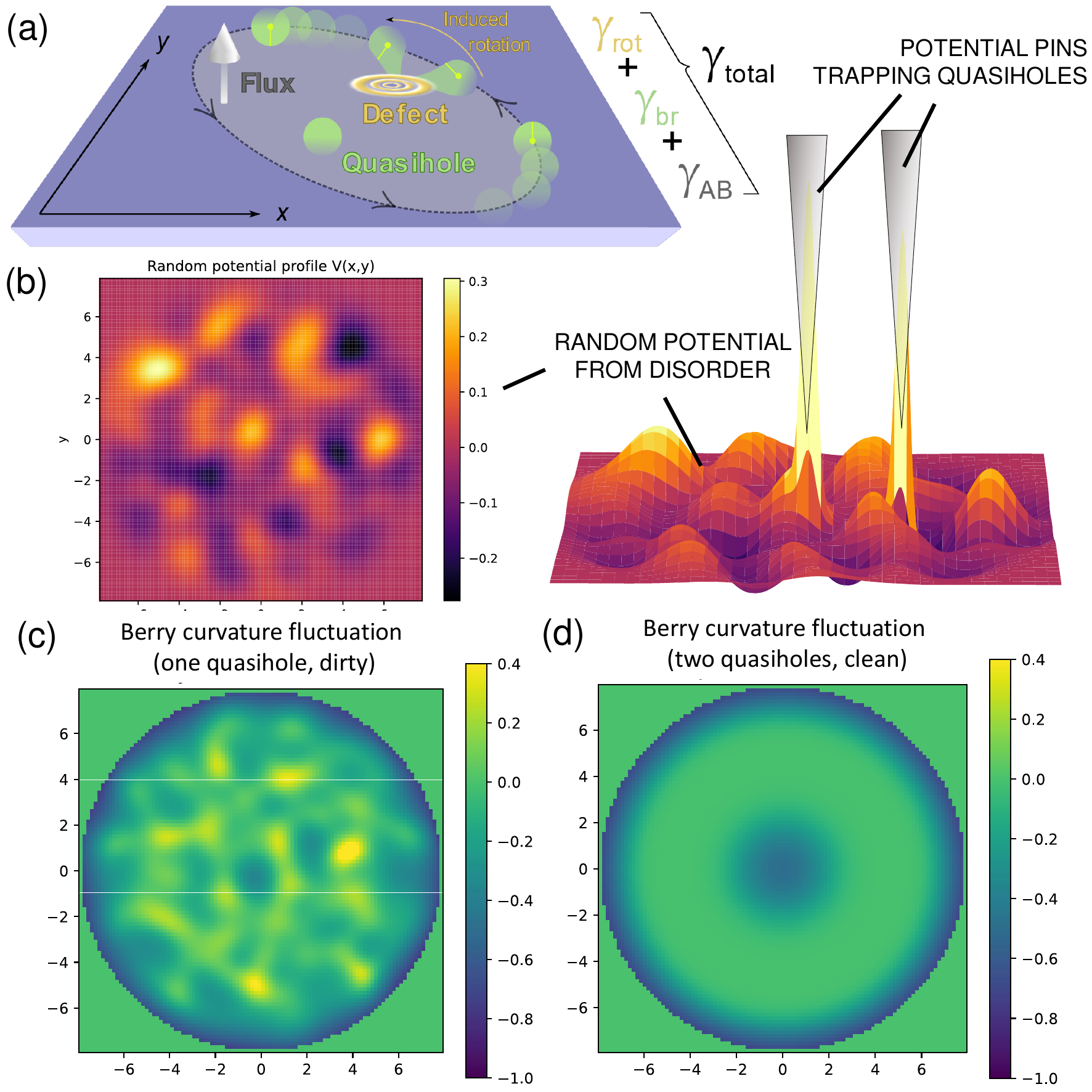}
\caption{(a) The total Berry phase consists of the AB phase, the statistical phase, and the self-rotation phase. The last term coming from a self-rotation induced by disorder near the path of the quasihole (b) Random potential profile $V(x,y)$ modelling a disordered system\cite{seesup}. The quasiholes are trapped and manipulated by two additional potential pins (e.g. an AFM tip) illustrated on the right panel (c) Deviation of local Berry curvature from average for a dirty system (d) Deviation of local Berry curvature from average for a clean system with an extra quasihole pinned at the center. Calculations done on a $\nu=1$ system with 30 electrons\cite{seesup}.}
\label{disorder}
\end{center}
\end{figure}

\textit{Effect of disorder--} 
The microscopic calculations in this work also allow us to study the effect of disorder on the Berry phase measurement in the bulk. Qualitatively, any disorder in the system can deform the shape of the quasiholes, therefore changing its intrinsic spin. When a quasihole moves past a region with disorder, a self-rotation is induced \emph{in general}, leading to an additional contribution to the Berry phase as described above. The effect of disorder on Berry phase anywhere on the QH droplet can be studied by calculating the local Berry curvature. We emphasize that our analysis applies to both quasiholes in the bulk and in the edge, as opposed to previous studies which focus quasiholes carried by the edge currents. In general, the Berry curvature contains three distinct distribution: the Aharanov-Bohm term from the background magnetic field, the presence of any extra quasiholes, and the presence of disorder (see Fig.\ref{disorder}a). 

While the Berry curvature is uniform in a clean system far from other quasiholes, local disorder induces fluctuation on the Berry curvature. If a quasihole travels along the edge, far away from the disorder in the bulk, the total Berry curvature fluctuation sums to zero as a result of Gauss-Bonnet theorem\cite{seesup}, consistent with the analysis in Ref. \cite{feldman2022robustness}. Any disorder at the edge can deform the quasihole shapes and thus modifies the Berry phase of those propagating along the edge, though the \emph{phase difference} when an addition or removal of another quasihole in the bulk will \emph{not be affected} by such disorder. 

However, procedures within the bulk may result in path-dependent Berry phases \emph{even if the loops encircle the same area and the same number of quasiholes}. This effect is attributed to the deformed internal structure of the quasihole and can be seen in the Berry phase fluctuation around the disorder points in Fig.\ref{disorder}c. This could be one of the main sources of noise in quantum computation from anyon braiding. 
Our analysis brings attention to the influence of disorder to the internal structures of quasiholes, and our method provides a tool to quantify such effects.

\textit{Conclusion and outlook--} 
In this paper, we have proposed the microscopic mechanism for the generalized spin-statistics for Abelian anyons in quantum Hall fluids. We have defined the intrinsic spin for quasiholes of the FQH phase from the algebra of the conformal Hilbert spaces (CHS), with the special case of identical quasihole clusters giving the standard spin-statistics relation. This intrinsic spin also applies to deformed quasiholes and can be detected from the proper Berry phase measurements. It should be noted while our discussion was limited to quasiholes in Abelian phases, the CHS algebra generalizes to Abelian quasielectrons and charged excitations of non-Abelian phases\cite{yang2021elementary,wang2021geometric}, and in principle, the intrinsic spin is well-defined for such cases. The hierarchy of CHS reveals the relationship between different FQH phases with interesting physical consequences; the implication of this on quasihole statistics remains to be studied, especially regarding non-Abelian anyons which are of interest to quantum computing. Our analysis poses questions about whether truly robust braiding in bulk is possible in a realistic system, as quasiholes are not point particles but objects with internal structures that can be influenced by disorder. The methods we proposed provide tools  to quantify such effects, which we explicitly illustrated with a new source of Berry phase from the local disorder deforming the shape of the quasiholes. In dirty systems where deviation from topological behavior is detected, our calculation can help to identify potential sources of errors and potentially devise methods to mitigate such effects.

\begin{acknowledgments}
We are grateful to L. Mazza and T. Comparin for helpful discussions. This work is supported by the NTU grant for Nanyang Assistant Professorship and the National Research Foundation, Singapore under the NRF fellowship award (NRF-NRFF12-2020-005), and a Nanyang Technological University start-up grant (NTU-SUG).
\end{acknowledgments}

\nocite{*}
\bibliographystyle{apsrev}
\bibliography{ref}

\clearpage
\renewcommand{\thefigure}{S\arabic{figure}}
\renewcommand{\thesection}{S\arabic{section}}
\renewcommand{\theequation}{S\arabic{equation}}
\renewcommand{\thepage}{S\arabic{page}}
\setcounter{figure}{0}
\setcounter{page}{1}
\setcounter{section}{1}

\onecolumngrid
\begin{center}
{\large \textbf{Online supplementary material for ``Spin-statistics relation and robustness of braiding phase for anyons in fractional quantum Hall effect"}}

\vspace{16pt}

\parbox{0.8\textwidth}{We provide detailed analysis leading to the proof that the Berry phase from rotating \emph{any} quasihole state by rotating potential traps equals $2\pi$ times the excess in angular momentum of the state compared to its rotationally invariant part. This is illustrated by using model wavefunctions on the sphere followed by a more general proof on the disk. This proof applies to all Abelian quasiholes including special case of the Moore-Read state. We also describe the numerical method for calculating the Berry phase of any physical procedure within a CHS, as well as for calculating the Berry curvature. This can be used to verify our results regarding the general spin-statistics relation. Extensive discussions concerning the internal structures of the quasiholes and their effects on the braiding phase is discussed. Finally we give a brief description of the composite fermionization process, which helps to numerically compute the intrinsic spin of the quasihole of any Abelian FQH state.}
\vspace{16pt}
\end{center}

\twocolumngrid

\section{Intrinsic angular momentum and braiding properties on the sphere}
\subsection{Rotating $k$-stack at the south pole - analytical derivation}
Here we provide more detailed calculation of the quaishole intrinsic angular momentum and braiding phase on the sphere. As discussed in the main text, by rotating a $k$-stack quasihole on the sphere one can extract its intrinsic angular momentum which couples to the curvature of the sphere. When there are two different stacked quasiholes, there is a further addition to the total Berry phase that is the braiding phase. This is calculation generalises the original arguments in Ref.\cite{li1992spin}.

On the sphere with a magnetic monopole of strength $2S$ placed at its center\cite{haldane1983fractional,greiter2011landau}, every particle on this system has angular momentum $L_z=-S,-S+1,...,S-1,S$, where $S$ is an integer due to Dirac quantization.  We define $L_z=S$ as the north pole (so $L_z=-S$ is the south pole). An FQH ground state is always a highest weight state at $L_z=L^2=0$. Quasiholes can then be added to this ground state by inserting additonal fluxes. In order to add localized quasihole that are $\hat L_z$ eigenstates, quasiholes are stacked at either the north pole or the south pole. Adding $k$ fluxes to the north pole (forming a $k$-stack quasihole at the north pole) decreases $L_z$ by $kN_e/2$ where $N_e$ is the number of electrons. An adiabatic rotation of the sphere about the z-axis by $2\pi$ thus give a Berry phase of $\gamma_1=-kN_e\pi$. On the other hand, if we create the $k-$stacked anyon at the south pole, the same argument applies and the Berry phase is $\gamma_2=kN_e\pi$.

$N_e$ can be related to the number of flux piercing the sphere, $N_\phi$, and other topological indices by
\begin{equation}
\label{condition}
N_{\phi}=\nu^{-1}N_e-s_c-s_f+k
\end{equation}
where (as discussed in the main text) $s_c$ is the cyclotron shift that depends on the Landau level (LL) index, and $s_f$ is the topological shift\cite{wen1992classification} that depends on the FQH phase. Thus the phase difference can be calculated accordingly
\begin{eqnarray}
\Delta\gamma_{21}=\gamma_2-\gamma_1&&=2\pi kN_e\\
&&=2\pi k\nu\left(N_{\theta}-k+s_f+s_c\right)\quad
\end{eqnarray}
which gives Eq.(3) in the main text.

When $(k+k')$ fluxes are added on the sphere, Eq.(\ref{condition}) becomes
\begin{equation}
\label{condition2}
N_{\phi}=\nu^{-1}N_e-s_c-s_f+k+k'
\end{equation}
We now consider two scenarios: one where the $(k+k')$-stack is at the north pole, which has $L_z=-(k+k')N_e/2$, and one where there is one $k'$-stack at the north pole and one $k$-stack at the south pole, which has $L_z=(k-k')N_e/2$. The difference between this two cases gives the total Berry phase when the $k$-stack encircles the $k'$-stack (on the disk this corresponds to the $k$-stack moving along an infinitely large circle with the $k'$-stack at the center). The phase difference is
\begin{eqnarray}
\Delta\gamma_{43} &&= \pi(k-k')N_e-\pi\left[-(k+k')N_e\right]\\
&&=2\pi kN_e\\
&&=2\pi\nu\left(N_\phi-k-k'+s_f+s_c\right)
\end{eqnarray}
which gives Eq.(7) in the main text.

\subsection{Rotating $k$-stack anywhere on the sphere - numerical result}
In first-quantized form, inserting a flux at position $z=a$ to a ground state wavefunction $\psi_{GS}(z)$ gives a single-flux state:
\begin{equation}
\label{one flux}
\psi_{1\phi}(z)\propto\prod_{i}(z_i-a)\psi_{GS}(z)
\end{equation}
where $z_i=x_i+iy_i$ is the holonomic variable parametrizing the position of the $i$-th electron. Similarly, inserting two fluxes at positions $a$ and $b$ gives
\begin{equation}
\label{two fluxes}
\psi_{2\phi}(z)\propto\prod_{i,j}(z_i-a)(z_j-b)\psi_{GS}(z)
\end{equation}
For numerical calculation, it is convenient to express these states in the basis that is the product state of $N$ particles in the LLL. The expansion of FQH ground states in this basis can be constructed from the Jack polynomial formalism\cite{bernevig2008model,bernevig2009clustering} -- in particular here we will consider the Laughlin and the Moore-Read states, which, respectively are Jack polynomials with root configuration $1001001...$ and $\alpha=-2$, and root configuration $1100110011...$ and $\alpha=-3/2$. Here we denote a many-body product state $|\phi_\lambda\rangle$ with a \emph{partition} $\lambda=(\lambda_1,\lambda_2,...,\lambda_N)$, $\lambda_i\in\mathbb{N}$, $\lambda_i<\lambda_i+1$ where:
\begin{equation}
\label{monomial state}
|\phi_\lambda\rangle\sim Asy\left\{|\lambda_1\rangle\otimes|\lambda_2\rangle\otimes...\otimes|\lambda_n\rangle\right\}
\end{equation}
where $|m\rangle$ denotes the LLL single-particle state with guiding center index $m$ and the operation $Asy\{\}$ anti-symmetrizes the $N$-particle state.

To determine the expansion of the quasihole states in Eq.(\ref{one flux}) and (\ref{two fluxes}) in this basis, a convenient result is the expansion of the prefactor:
\begin{equation}
\label{factor}
\prod_i(z_i-a) = \sum_{k=0}^{N}(-a)^ke_{N-k}(z)
\end{equation}
where $e_k(z)$ is the \emph{elementary symmetric monomial} of order $k$. The result of multiplying an $e_k(z)$ to a many-body product state is known,
\begin{equation}
\label{e_k}
e_k(z)\phi_{\lambda}(z)=\phi_{\lambda^{(1)}_{[k]}}(z)+\phi_{\lambda^{(2)}_{[k]}}(z)+...
\end{equation}
where each $\lambda^{(i)}_{[k]}$ is obtained by picking some $k$ elements in $\lambda$ and increasing it by one. There are at most ${N\choose k}$ such terms, but some terms may vanish if the resulting partition has two or more equal parts (which is forbidden by the Pauli exclusion principle for electrons).
Eqs. (\ref{factor}) and (\ref{e_k}) can be used to contruct the many-body state expansion for the states in Eqs.(\ref{one flux}) and (\ref{two fluxes}) by a recursive numerical routine. 

Here we are interested in the state living on the sphere, and hence the states must be normalized accordingly. On the sphere the guiding center index $m$ can take $2S+1$ different values: $-S,-S+1,...,S-1,S$, which is the same as the $L_z$ eigenvalues described above. The position of a particle is parametrized by the azimuthal angle $\theta$ and polar angle $\varphi$, and the holonomic variable $z$ is related by the stereographic projection\cite{jain2007composite} 
\begin{equation}
\label{stereographic}
z=\tan{\left(\theta/2\right)}e^{i\varphi}
\end{equation}
One can verify that this gives us a definition of the inner product between any pair of basis state as
\begin{equation}
\label{inner product}
\langle\phi_{\lambda}|\phi_{\mu}\rangle\equiv\int d\Omega\frac{\phi_{\lambda}^*\phi_{\mu}}{c_\lambda c_\mu}
\end{equation}
where
\begin{equation}
\label{sphere coefficient}
c_\lambda = \prod_{i=1}^N\sqrt{\frac{(2S)!}{(S-\lambda_i)!(S+\lambda_i)!}}
\end{equation}

With the Moore-Read state, the situation gets a little more complicated. Adding a flux tothe Moore-Read ground state yields a quasihole state:
\begin{equation}
\label{MR quasihole}
\psi(z) = \prod_i(z_i-a)Pf\left(\frac{1}{z_i-z_j}\right)\prod_{i<j}(z_i-z_j)^2
\end{equation}
(where $Pf()$ denotes the Pfaffian). For the Moore-Read state, a single flux can be split into two half-fluxes, each of which acts as a separate quasiholes. The first-quantized wavefunction for the general two-half-flux state is
\begin{widetext}
\begin{equation}
\label{MR quasiholes split}
\psi(z) = Pf\left(\frac{(x_i-a_1)(x_j-a_2)-(x_i-a_2)(x_j-a_1)}{z_i-z_j}\right)\prod_{i<j}(z_i-z_j)^2
\end{equation}
\end{widetext}
where the two quasiholes are positioned at $a_1$ and $a_2$. Unlike the wavefunctions of the form Eq.(\ref{one flux}) and (\ref{MR quasihole}), the expansion coefficient of Eq.(\ref{MR quasiholes split}) in the monomial basis is not known. Therefore, we resolve to numerical construction of states with two localized half-fluxes by taking the ground state of local trapping potential. On the sphere, a one-body potential centered at a point $(\theta_0,\phi_0)$, where $\theta_0$ and $\phi_0$ are respectively the azimuthal and polar angles, can be constructed as
\begin{equation}
\label{V 1body}
\hat V_{1bdy} = |\theta_0,\phi_0\rangle\langle\theta_0,\phi_0|
\end{equation}
where $|\theta_0,\phi_0\rangle$ is the single-particle coherent state centered at $(\theta_0,\phi_0)$, constructed by applying the magnetic rotation operator (the spherical analog of magnetic translation operator on the plane) on the coherent state centered at the north pole. Given the single-particle angular momentum basis $\{|\phi_m\rangle, m=0,1,...,2S\}$, the matrix element of this one body operator can be calculated as
\begin{equation}
\label{V 1body matrix}
V_{m,n} = \langle\phi_m|\hat V|\phi_n\rangle = \bar c_m c_n \bar\phi_m(z)\phi_n(z)
\end{equation}
where $c_m$ is the coefficient of $|\phi_m\rangle$ in the angular momentum basis expansion of $|\theta_0,\phi_0\rangle$. This matrix element in the single-particle basis can in turn be used to calculate the matrix element in the manner described in Section S5.C.


\section{Trapping and moving quasiholes with a pinning potential: Berry phase derivation on the disk}
\subsection{Perturbative method}
In an idealized scenario, a quasihole can be localized by applying a potential trap on an electron gas at a certain FQH phase. Any physical operation on the potential trap will affect the quasihole accordingly. Here we consider a quasihole rotating about its own center, which can be realized by rotating the potential trap. This rotation is only physically meaningful if the potential profile is not rotationally invariant. We can write the potential generally as
\begin{equation}
\label{trapping potential}
\hat H(\theta)=\hat H_0+\lambda \hat H_1(\theta)
\end{equation}
where $\hat H_0$ is a rotationally invariant function (e.g. a Dirac delta function or a cylindrical well) and $\lambda \hat H_1$ with $|\lambda|\ll1$ is a small perturbation to break (continuous) rotationally symmetry. The Hamiltonian is parametrized with $\theta$ which denotes its orientation on the $x$-$y$ plane, and we can rotate the Hamiltonian by tuning $\theta$ from 0 to $2\pi$. We consider the result of this rotation on the ground state:
\begin{equation}
\label{ground state}
\hat H(\theta)|\psi^{(0)}(\theta)\rangle=E^{(0)}|\psi^{(0)}(\theta)\rangle
\end{equation}
This rotation gives a Berry phase:
\begin{equation}
\label{Berry phase}
\gamma = i\int \langle\psi(\theta)|\frac{d}{d\theta}|\psi(\theta)\rangle d\theta
\end{equation}
where by defining the Berry connection
\begin{equation}
\label{Berry connection}
A_\theta = i \langle\psi(\theta)|\frac{d}{d\theta}|\psi(\theta)\rangle
\end{equation}
one may note the symmetry of the problem and expect the total phase to be $2\pi A_\theta$ (i.e. the Berry connection is independent of $\theta$). Below we will calculate this Berry connection. Throughout this section we will assume the groundstates of both $\hat H_0$ and $\hat H(\theta)$ are non-degenerate.

Let $|\psi_0^{(n)}\rangle$ be the $n$-th eigenstate of $\hat H_0$:
\begin{equation}
\hat H_0|\psi_0^{(n)}\rangle=E_0^{(n)}|\psi_0^{(n)}\rangle
\end{equation}
The ground state of $\hat H_0$ can be written as
\begin{align}
|\psi^{(0)}\rangle &=|\psi_0^{(0)}\rangle + \lambda |\psi_1^{(0)}\rangle\label{ground state 1}\\
|\psi_1^{(0)}\rangle &=\sum_{n>0}{\frac{|\psi_0^{(n)}\rangle\langle\psi^{(n)}_0|\hat H_1|\psi_0^{(0)}\rangle}{E_0^{(n)}-E_0^{(0)}}}\label{ground state 2}
 \end{align}
here $|\lambda_1^{(0)}\rangle=|\lambda_1^{(0)}(\theta)\rangle$ inherit the $\theta$-dependence from $\hat H_1$. Since $|\psi_0\rangle$ is independent of $\theta$, the differential $d/d\theta$ in Eq.(\ref{Berry connection}) only acts on $|\psi_1\rangle$. To see its result, consider an infinitesimal rotation by $\delta\theta$ is described by the unitary operator $e^{i\delta\theta\hat L_z}$, where $\hat L_z$ is the angular momentum operator. $\hat H_1$ is rotated as
\begin{align}
\hat H_1(\theta+\delta\theta) &= e^{i\delta\theta \hat L_z}\hat H_1(\theta)e^{-i\delta\theta\hat L_z}\label{rotate potential 1}\\
&= \hat H_1(\theta)+i\delta\theta\left[\hat L_z, \hat H_1(\theta)\right]\label{rotate potential 2} + \mathcal O(\delta\theta^2)
\end{align}
As a result one finds
\begin{equation}
\label{d psi 1}
\frac{d}{d\theta}|\psi_1^{(0)}\rangle=i\sum_{n>0}{\frac{|\psi_0^{(n)}\rangle\langle\psi^{(n)}_0|\left[\hat L_z ,\hat H_1\right]|\psi_0^{(0)}\rangle}{E_0^{(n)}-E_0^{(0)}}}
\end{equation}
Note that since $\hat H_0$ commutes with $\hat L_z$ (by definition of rotational invariance), every eigenstate of $\hat H_0$ is also an eigenstate of $L_z$:
\begin{equation}
\label{Lz eigenstate}
\hat L_z|\psi^{(n)}_0\rangle=\ell_n|\psi_0^{(n)}\rangle
\end{equation}
Thus Eq.(\ref{d psi 1}) simplifies to
\begin{equation}
\frac{d}{d\theta}|\psi_1^{(0)}\rangle=i\sum_{n>0}{\frac{(\ell_n-\ell_0)|\psi_0^{(n)}\rangle\langle\psi^{(n)}_0|\hat H_1|\psi_0^{(0)}\rangle}{E_0^{(n)}-E_0^{(0)}}}
\end{equation}
which gives
\begin{align}
A_\theta &=i|\lambda|^2 \langle\psi_1^{(0)}|\frac{d}{d\theta}|\psi_1^{(0)}\rangle\\
&= -|\lambda|^2\left(\langle\psi_1^{(0)}|\hat L_z|\psi_1^{(0)}\rangle - \ell_0\right)
\end{align}

On the other hand, calculating the average angular momentum of Eq.(\ref{ground state 1}) yields (to the second order in $\lambda$):
\begin{align}
\langle\psi^{(0)}|\hat L_z|\psi^{(0)}\rangle &= \langle\psi_0^{(0)}|\hat L_z|\psi_0^{(0)}\rangle + |\lambda|^2\langle\psi_1^{(0)}|\hat L_z|\psi_1^{(0)}\rangle\label{Lz 1}\\
&=\ell_0 +|\lambda|^2\langle\psi_1^{(0)}|\hat L_z|\psi_1^{(0)}\rangle\label{Lz 1}
\end{align}
Thus we get:
\begin{equation}
A_\theta = -\left( \langle\psi^{(0)}|\hat L_z|\psi^{(0)}\rangle -  \langle\psi_0^{(0)}|\hat L_z|\psi_0^{(0)}\rangle\right)
\end{equation}
In other words, the self-rotation phase of a quasihole state is proportional to its excess of angular momentum compared to the rotationally invariant part. Physically this reflects the fact that a perfectly symmetric quasihole cannot be rotated (the Berry connection vanishes as $\theta\to0$). To observe a nontrivial self-rotational phase, a small perturbation that breaks rotational symmetry has to be added. This results in an additional angular momentum that can be detected as Berry phase.

\subsection{Nonperturbative method}
In the main text we argue that the quasihole self-rotation discussed above generalizes to rotation of any quasihole state. In particular in the case of multiple quasihole it gives rise to an additional contribution from the quasihole statistics. To show this rigorously, in this section we generalize the above discussion to the case with finite $\lambda$.

When the addition to $\hat H_0$ is nonperturbative, the eigenstates of $\hat H_0$ still forms a complete orthonormal basis for the Hilbert space. We can thus expand the ground state of $\hat H$ as
\begin{equation}
\label{ground state general}
|\psi^{(0)}\rangle = \lambda_0|\psi^{(0)}_0\rangle + \sum_{n>0}\lambda_n|\psi_0^{(n)}\rangle\text{, for } \lambda_0\in\mathbb{R}
\end{equation}
where $\sum_{i\geq 0}|\lambda_i|^2=1$ normalizes the state. Here the restriction that $\lambda_0$ is a real number serves to fixes the gauge of the ground state. Without this gauge fixing, any state of the form $e^{i\phi(\theta)}|\psi^{(0)}\rangle$ for some real function $\phi(\theta)$ would be an equally valid ground state of $\hat H$, but this would add an additional contribution to the Berry connection that is ultimately non-physical\cite{berry1984quantal}. Note that in the previous section this restriction was implicitly implemented as the coefficient of $|\psi_0^{(0)}\rangle$ in Eq.(\ref{ground state 1}) is always unity.

\begin{figure*}
\begin{center}
\includegraphics[width=0.8\linewidth]{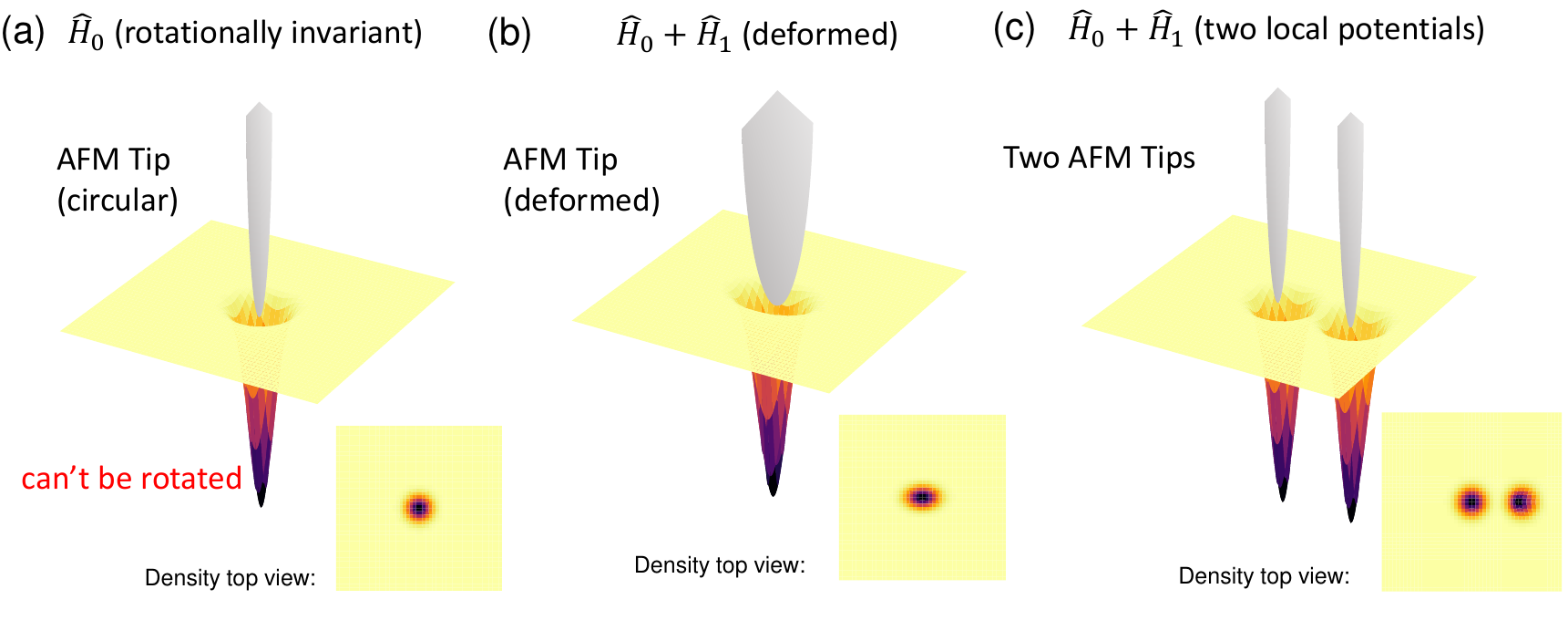}
\caption{(a) A quasihole (represented by a dip in electron density in both the 3D plot and the inset heatmap) trapped with a perfectly rotationally symmetric potential (e.g. with a perfect AFM tip) cannot be physically rotated (b) For rotation to be physically meaningful, the potential must be deformed (e.g. using an AFM tip with an elliptical cross-section). This deformation also modifies the quasihole spin. (c) A second AFM tip plays the same role mathematically as the deformation of the first AFM tip. This perspective unifies the exchange statistics and quasihole spin into a single microscopic process.}
\label{fig:potentials}
\end{center}
\end{figure*}

An infinitesimal rotation to $\hat H$ results in $\hat H(\theta+\delta\theta)=e^{i\delta\theta\hat L_z}\hat H(\theta)e^{-i\delta\theta\hat L_z}$. One can easily check that the ground state of this Hamiltonian is given by the ansatz
\begin{equation}
\label{rotate ground state 3}
|\psi^{(0)}(\theta+\delta\theta)\rangle = e^{i\delta\theta\left(\hat L_z-\ell_0\right)}|\psi^{(0)}(\theta)\rangle
\end{equation}
which can also be expanded as Eq.(\ref{ground state general}). We can then write
\begin{equation}
\label{differential}
\frac{d}{d\theta}|\psi^{(0)}\rangle = i\sum_{n>0}\lambda_n \ell_n|\psi_0^{(n)}\rangle
\end{equation}
(note that here Eq.(\ref{Lz eigenstate}) still applies). Thus the Berry connection can be calculated as
\begin{align}
A_\theta&=i\langle\psi^{(0)}|\frac{d}{d\theta}|\psi^{(0)}\rangle\label{general Berry connection 1}\\
&= -\sum_{n>0}|\lambda_n|^2(\ell_n-\ell_0)\label{general Berry connection 2}\\
&= -\left( \langle\psi^{(0)}|\hat L_z|\psi^{(0)}\rangle -  \langle\psi_0^{(0)}|\hat L_z|\psi_0^{(0)}\rangle\right)\label{general Berry connection 3}
\end{align}
Thus, we have proven that the Berry phase obtained by rotating \emph{any} state is equal to the excess of its angular momentum compared to the rotationally invariant part. As discussed in the main text, this is the physical process that describes the Berry phase on the plane which is the stereographic counterpart of the derivation on the sphere. This in turn gives us the microscopic origin of the general spin-statistics relation, since having multiple quasiholes is just a special case of Eq.(\ref{trapping potential}), with $\hat H_1$ being a second potential trap at a separate location (see Fig.\ref{fig:potentials}) .

\subsection{Equivalence between the sphere and disk calculations}
The above analysis offers two notable advantages compared to the calculation on the sphere. Firstly, while the sphere analysis makes use of ansatz states and hence is limited to model states, the disk analysis can be generalized to quasiholes with any shape. That is because we have not made any restriction on the pinning Hamiltonian in Eq.(\ref{trapping potential}), only that $\hat H_0$ be rotationally invariant (the total $\hat H$ can take any shape and form). Secondly, despite a seemingly a different starting point, the disk analysis is a generalization of the sphere calculation, which we show in this section.

A gaussian wavepacket centered at the center of the disk is the zero-eigenvalue coherent state of both the cyclotron and guiding center ladder operators: $\hat a|0\rangle=\hat b|0\rangle=0$. This coherent state can be translated to be centered at any position $\mathbf{r}$ by the magnetic translation operator:
\begin{equation}
\label{coherent state}
|\mathbf{r}\rangle = \hat T_{\mathbf r}|0\rangle
\end{equation}
This state can be used to construct a local Hamiltonian as follows:
\begin{equation}
\label{potential}
\hat V(\mathbf r) = V_0|\mathbf r\rangle\langle \mathbf r|
\end{equation}
which in real space gives the shape of a gaussian potential bump. This gives the single-particle matrix element $V_{mn}=\langle m|\hat V|n\rangle$, which can then be used to construct the many-body matrix representation (see the next section). It is easy to check that the one-quasihole state of the same form as Eq.(\ref{one flux}), where $a=r_x+ir_y$ and $\psi_{GS}$ is the ground state of any FQH phase, is the exact zero state of this one-body potential, and the two-quasihole state in Eq.(\ref{two fluxes}) is the exact zero state of a potential profile consisting of two distinct Gaussian packet. Eq.(\ref{one flux}) and (\ref{two fluxes}) in turn provides a stereographic mapping between the disk and the sphere. Thus we see that the two separate derivation on two different geometries are actually equivalent. Eq.(\ref{general Berry connection 3}) however is more general since it applies to any quasihole state, not just model quasihole states taking the form of Eq.(\ref{one flux}) or (\ref{two fluxes}).

\section{Hollow-core quasihole}
We discuss here a particularly interesting deformation to a quasihole cluster that changes the intrinsic spin without altering the exchange phase. We call this type of quasihole  ``hollow-core" (see Fig.\ref{fig:hollow core}b). On the sphere, a hollow-core quasihole can be obtained by placing a $k$-stack at the north pole and applying the $\hat L^-$ operator on the state. This operation pulls one quasihole away from the center of the stack. We denote this species of hollow-core quasihole with $k\setminus 1$.
\subsection{One quasihole}
The intrinsic angular momentum of a hollow-core quasihole can be calculated on the sphere in the same manner as for the $k$-stack in the main text. The Berry phase is given by $2\pi$ times the difference in angular momenta of the states with the quasihole at the south pole and the north pole. At the north pole, a $k$-stack quasihole has angular momentum $kN_e/2$ where $kN_e$ is the number of electrons. A hollow core quasihole is created by applying the $\hat L^-$ operator on this state; thus the angular momentum of the state is $kN_e/2-1$. Similarly, the hollow core at the south pole has angular momentum $-kN_e/2+1$. The total Berry phase is
\begin{align}
\gamma_1 &= 2\pi\left(\langle L_z\rangle_{south}-\langle L_z\rangle_{north}\right)\label{berry1}\\
&=4\pi(kN_e/2-1)\label{berry2}
\end{align}
On the sphere $N_e$ is related to the number of fluxes $N_\phi$ as
\begin{equation}
\label{flux}
k+\nu^{-1}N_e-s_f = N_\phi+s_c
\end{equation}
where $s_f$ and $s_c$ are respectively the topological shift and cyclotron shift as discussed in the main text. Substituting this into Eq.(\ref{berry2}), one can see that the intrinsic angular momentum is
\begin{equation}
\label{hollow core am}
s_{k\setminus1} =-\frac{\nu k^2}2 - 1 + \frac{\nu s_fk}2 + \frac{\nu s_c k}2
\end{equation}
Thus the topological spin can be read off as
\begin{equation}
\label{hollow core topo spin}
s^{topo}_{k\setminus1} = -\frac{\nu k^2}2 -1
\end{equation}

\subsection{Two quasiholes}
Similarly, the total Berry phase for braiding two hollow-core quasiholes can be taken to be $2\pi$ times the difference in angular momentum of two states. The first is with one quasihole at each pole (which on the disk corresponds to braiding one quasiholes along an infinitely large circle around the other), and the second is with both quasihole stacked on the north pole. Here there is an ambiguity in choosing this second state. Intuitively, one might view stacking two $k\setminus1$ quasiholes as a single $2k\setminus2$ quasiholes. However, the $2k$-stack is also a possible choice, as it corresponds to the ground state of a single rotationally invariant local potential $H_0$. However, we see that the angular momenta of these two choices differ by 2 ($2k\setminus2$ quasihole is obtained from a $2k$-stack by applying the $L^-$ operator twice). This corresponds to a difference of $4\pi$ in the calculated phase, which has no real physical significance.

\begin{figure}
\begin{center}
\includegraphics[width=\linewidth]{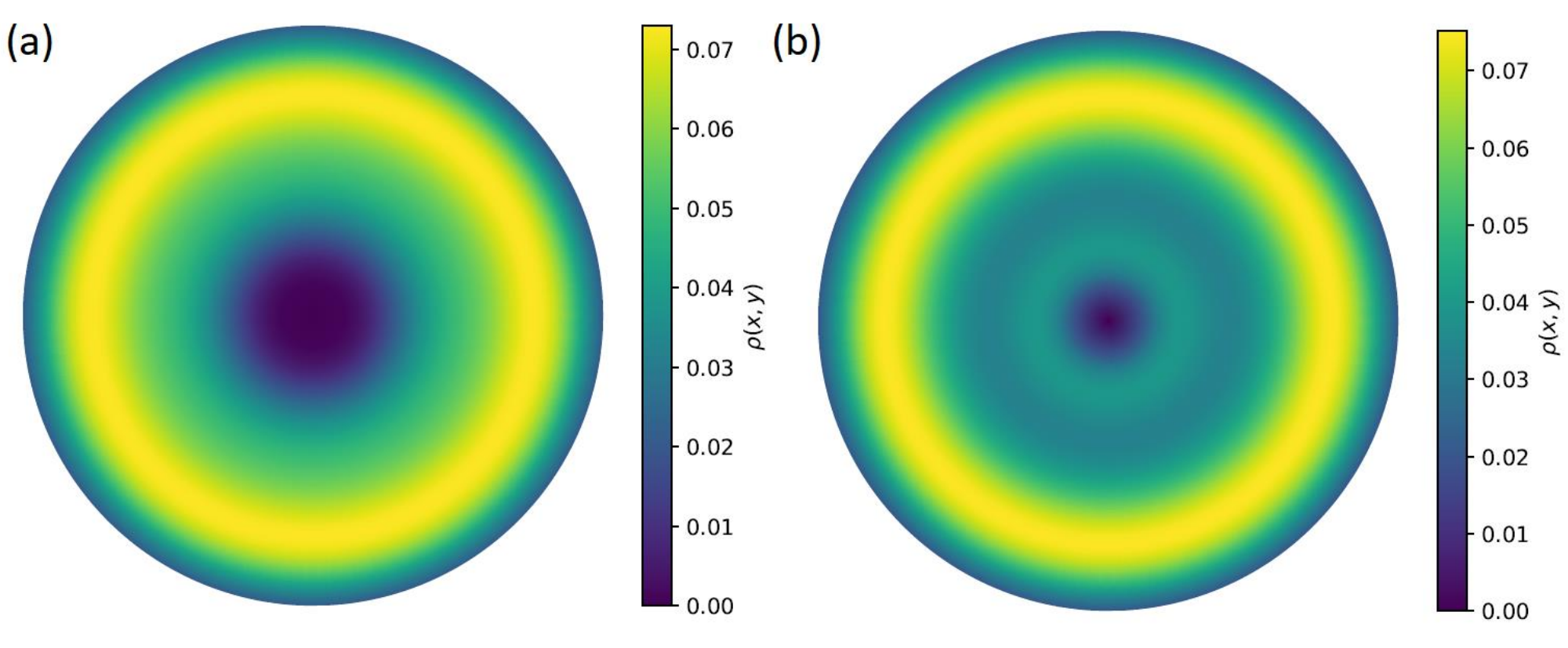}
\caption{The quasihole of a Laughlin state at $\nu=1/3$. The quasihole manifests as a local region of deficiency in electron density (dark blue region around the center) (a) 2-stack quasihole (Jack polynomial with root configuration $001001001...$) (b) $2\setminus1$ hollow-core quasihole (Jack polynomial with root configuration $010001001001...$) }
\label{fig:hollow core}
\end{center}
\end{figure}

Taking the $2k\setminus2$ quasihole as the north pole stack, we find that the total Berry phase is
\begin{equation}
\label{berry 2qh}
\gamma_2 = -4\pi(kN_e/2-1)
\end{equation}
which is of similar form to Eq.(\ref{berry2}), but here $N_e$ and $N_\phi$ are related as
\begin{equation}
2k+\nu^{-1}N_e-s_f = N_\phi+s_c
\end{equation}
Taking the difference in the total phase, we find that the braiding phase is
\begin{equation}
\label{hollow core braid}
\gamma_{br}\equiv \gamma_2-\gamma_1 = 2\pi\nu q^2
\end{equation}
which is the same as braiding two regular $k$-stacks. This is rather unsurprising and demonstrate the argument that the braiding phase of two quasiholes comes from the deficiency in electron density within the enclosed area\cite{arovas1984fractional,einarsson1995fractional}: both the $k$-stack and the hollow core quasiholes result in the same amount of deficiency, only with different shapes (see Fig.\ref{fig:hollow core}). However, comparing this braiding phase with the topological spin in Eq.(\ref{hollow core topo spin}), we emphasize that the general spin-statistics theorem discussed in the main text gives the right relation between the two quantities.

\section{Numerical method}
\subsection{Matrix elements of a single particle $\delta$-potential on the LLL}
Here we present an example of how to write down the matrix element of a given potential in an isotropic system, which means that we have the angular momentum as a good quantum number to label the single-particle orbits $| n \rangle$, so the density operator matrix element in this basis is given by:
\begin{equation}
\langle m | \hat{\rho}_{\boldsymbol{q}} | n \rangle = \sqrt{\frac{m!}{n!}} ( i \cdot \tilde{\boldsymbol{q}})^{n-m} L_m^{(n-m)} \left(\tilde{Q} \right) e^{-\frac{1}{2} \tilde{Q}}
\end{equation}
and we can adopt the LLL form factor without loss of generality:
\begin{equation}
F_0(\boldsymbol{q}) = \langle M | \hat{\tilde{\rho}}_{\boldsymbol{q}} | N \rangle = \langle 0 | \hat{\tilde{\rho}}_{\boldsymbol{q}} | 0 \rangle =  e^{-\frac{1}{2} \tilde{Q}}
\end{equation}
where we have defined:
\begin{equation}
\tilde{\boldsymbol{q}} = \frac{1}{\sqrt{2}} (q_x - i q_y); \quad \tilde{Q} \equiv |\tilde{\boldsymbol{q}}|^2 = \frac{|\boldsymbol{q}|^2}{2}
\end{equation}


For a single $\delta$-potential at $(x_0, 0)$, the unitary Fourier transform with angular frequency in polar coordinates is given by the translation of $\tilde{V}_{0}(r_q, \theta_q)$:
\begin{equation}
\begin{aligned}
V_{\boldsymbol{q}} = \tilde{V}_{x_0}(r_q, \theta_q) &=  e^{i x_0 \cdot q_x} = e^{i x_0 \sqrt{2 \tilde{Q}}   \cos(\theta_{q})}
\end{aligned}
\end{equation}

Then the matrix element can be written as:
\begin{small}
\begin{equation}
\begin{aligned}
H^{x_0}_{m n} = & \int d^2 \boldsymbol{q}  \langle m | \hat{\rho}_{\boldsymbol{q}} | n \rangle F_0(\boldsymbol{q}) \cdot V_{\boldsymbol{q}}\\
= & \pi \sqrt{\frac{m!}{n!}} \cdot \int_0^{\infty}  d \tilde{Q} \cdot i^{n-m}\\
& \cdot J_{n-m}\left( x_0 \sqrt{2 \tilde{Q}} \right) \cdot   \left(i \cdot \sqrt{ \tilde{Q}} \right)^{n-m} L_m^{(n-m)} \left(\tilde{Q} \right) e^{- \tilde{Q}}
\label{matrixelementint1}
\end{aligned}
\end{equation}
\end{small}
where $J_{\alpha}(x)$ is the Bessel function of the first kind, which is related to the Laguerre polynomials by:
\begin{equation}
\begin{aligned}
J_{n-m}\left( x_0 \sqrt{2 \tilde{Q}} \right) = &\left(\frac{ x_0 \sqrt{ \tilde{Q}}}{\sqrt{2}}\right)^{n-m} \cdot \frac{e^{-\frac{x_0^2}{2}}}{\Gamma(n-m+1) }\\
&\cdot  \sum_{k=0}^{\infty} \frac{L_{k}^{(n-m)}\left(\tilde{Q}\right)}{\left(\begin{array}{c}
k+n-m\\
k
\end{array}\right)} \frac{\left(\frac{x_0^2}{2}\right)^{k}}{k !}
\label{bess_lag}
\end{aligned}
\end{equation}
substituting which into Eq.\ref{matrixelementint1} gives:
\begin{small}
\begin{equation}
\begin{aligned}
H^{x_0}_{n \ge m}
&=   \frac{(-1)^{n-m}\cdot \pi}{\sqrt{2}^{m+n} \cdot \sqrt{m! \cdot n!}} \cdot  x_0^{m+n} e^{-\frac{x_0^2}{2}}
\label{matrixelementint2}
\end{aligned}
\end{equation}
\end{small}

One can easily see the symmetry between $x$- and $y$-direction, so for the $\delta$-potential at $(x_0, y_0)$, the matrix elements of the Hamiltonian are given by:
\begin{equation}
\label{delta matrix}
H_{m n}(\boldsymbol{R}) =
\frac{ \pi}{\sqrt{m! \cdot n!}} \cdot  \left(\frac{R}{\sqrt{2}}\right)^{m+n} e^{-\frac{R^2}{2}} e^{i(n-m) \cdot \theta_R}
\end{equation}
where $\boldsymbol{R} \equiv R \cdot e^{i \theta_R} = x_0 + i \cdot y_0$.

In fact the eigenstates and the eigenvalues of this Hamiltonian can be \textit{rigorously} solved. However $V_q$ can be any function (commonly-used ones including $\nabla^{2} \delta^{(2)}(\boldsymbol{r})$, cylindrical functions, Gaussian functions, etc.) so the corresponding Hamiltonian could be very complicated.

\subsection{Matrix-element of one-body local potential}
In a more general case, it is usually not possible to derive an analytical formula for the matrix element of a general one-body potential. In that case, the matrix element can be calculated by numerical integration: given a one-body potential profile $V(z)$, its matrix representation in the monomial basis can be calculated as
\begin{equation}
\label{potential matrix}
V_{mn}=\int d^2z V(z)\phi_m^*(z)\phi_n(z)
\end{equation}
where $\phi_m(z)\propto z^me^{-|z|^2/4l_B^2}$ is the single-particle state on the disk geometry. The integral is calculated over the entire disk with radius $R=\sqrt{2N_o}l_B$ where $N_o$ is the number of orbitals in the finite system. For all numerical routines we set the magnetic length $l_B=1/\sqrt{eB}$ to 1 (as a result, all length quantities shown are in the unit of magnetic length.) 

\begin{figure}
\begin{center}
\includegraphics[width=\linewidth]{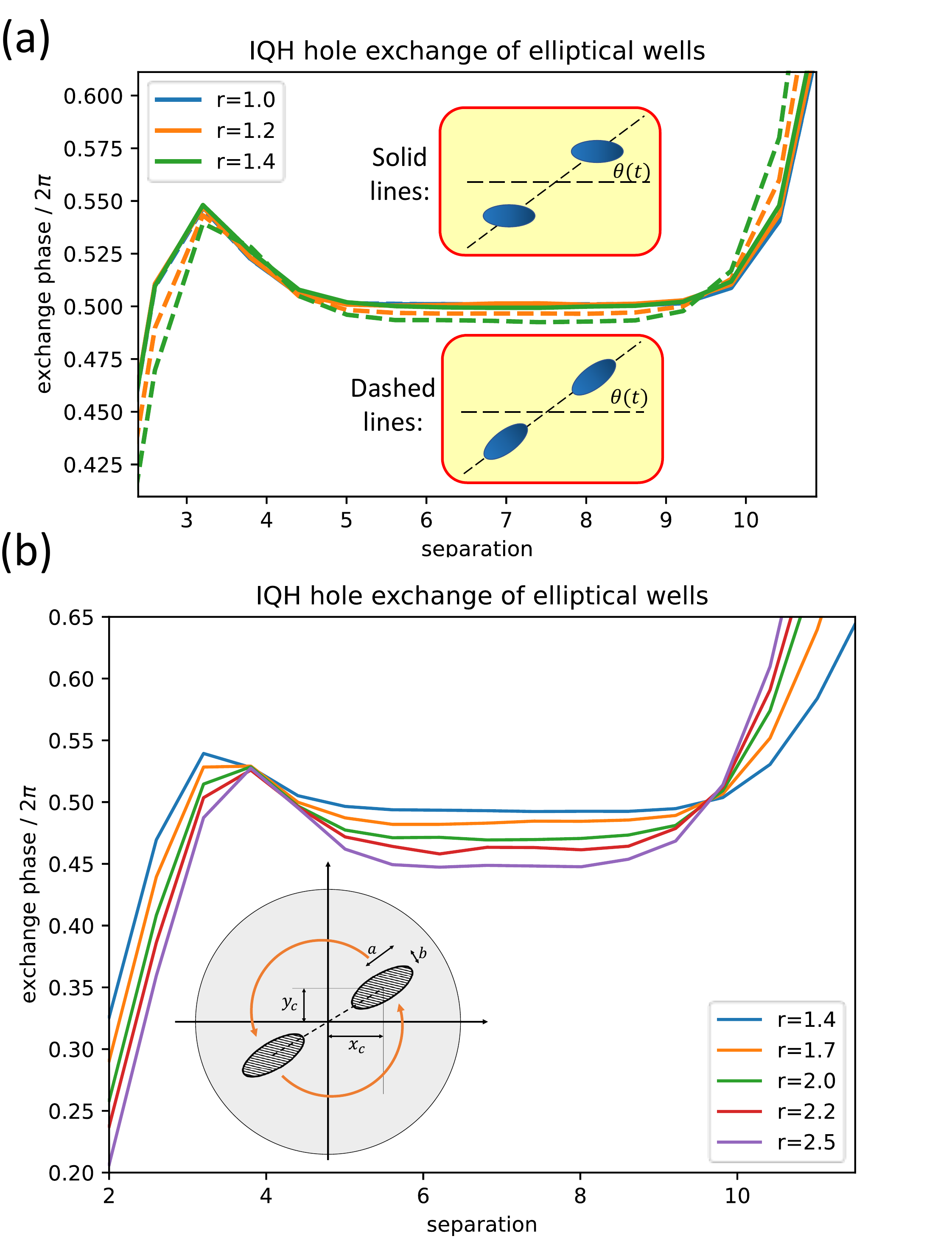}
\caption{(a) Insets shows two different exchange/braiding schemes when the quasiholes are not rotationally invariant. Exchanging squeezed IQH holes by pure translation (solid lines) preserves the fermionic statistics while exchanging with rotation (dashed lines) shows deviation. (b) Exchanging elliptical holes (IQH state at $\nu=1$) with rotation shows deviation from the fermionic statistics. Inset shows the schematic of the setup on the QH droplet with the shaded regions showing the elliptical wells. All numerics are done on an IQH system with $N=24$ electrons.}
\label{fig:rotate ellipse}
\end{center}
\end{figure}

\begin{figure*}
\begin{center}
\includegraphics[width=\linewidth]{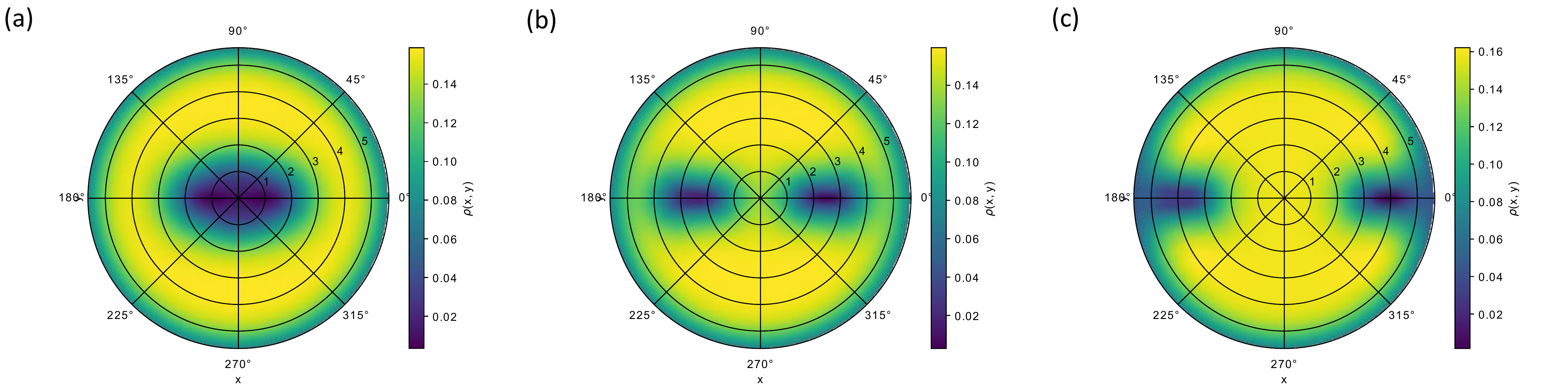}
\caption{Density of two squeezed IQH holes with squeezing ratio $r=2$ at different separation distances: (a) $2|r_0|=2$ (b) $2|r_0|=5$ (c) $2|r_0|=8$}
\label{fig: ellipse density}
\end{center}
\end{figure*}

In particular, an elliptical potential pin can be constructed as:
\begin{equation}
V^{ellipse}_{r_0, R}(\mathbf{r}) =
\begin{cases}
V_0 & d^ad^bg_{ab}<R, \mathbf d\equiv \mathbf r - \mathbf r_0\\
0 & \text{otherwise}
\end{cases}
\end{equation}
which is non-zero only within an elliptical region with area $\pi R^2$ centered at $\mathbf r_0=(x_0,y_0)$. The shape of the ellipse is parametrized by a unimodular metric $g_{ab}$:
\begin{equation}
g =
\begin{pmatrix}
\cosh{\theta}+\sinh{\theta}\cos{\phi} & \sinh{\theta}\sin{\phi}\\
\sinh{\theta}\sin{\phi} & \cosh{\theta}-\sinh{\theta}\cos{\phi}
\end{pmatrix}
\end{equation}
Here $\theta$ parametrizes the squeezing ratio and $\phi$ parametrizes the rotation w.r.t. the $x$-axis. This potential can be used to numerically study the self-rotational phase of one quasihole and the exchange property of two quasiholes. In the case of exchanging two quasiholes, we construct a Hamiltonian
\begin{equation}
\begin{split}
H(R,\theta) &= V^{ellipse}_{(-x_0,-y_0),R} + V^{ellipse}_{(x_0,y_0),R}\\
x_0&=R_0 \cos{\theta}\\
y_0&=R_0 \sin{\theta}
\end{split}
\end{equation}
and tune $\theta$ between $(0,\pi)$. The Berry phase is calculated as Eq.(\ref{cumulative overlap}). In general, the exchange statistics is obtained at large separation distance: $2|r_0|\to\infty$ where $|r_0|=\sqrt{x_0^2+y_0^2}$. However on a finite system this is not possible and one must ensure the two quasiholes are well-separated without falling off the edge of the quantum Hall droplet (Fig.\ref{fig: ellipse density}c). This drastically limits the possible cases that can be studied numerically even for simple states such as the Laughlin 1/3 state. However we claim that the results from the study on the IQH state at $\nu=1$ shown below generalize to all Abelian FQH states.

\subsection{Matrix element of one-body potential in many-body basis}
Given the matrix elements of a one-body potential in the single-particle basis, such as in Eq.(\ref{delta matrix}) or (\ref{potential matrix}), one can rewrite the potential operator in second-quantized form:
\begin{equation}
    \label{one-body potential}
    \hat V^{1bdy} = \sum_{m,n}V_{mn}\hat c^\dagger_m\hat c_n
\end{equation}
where $\hat c^\dagger_m$ and $\hat c_n$ respectively creates and destroy an electrons at orbital indexed $m$ and $n$. We can then calculate the matrix element of $\hat V$ given a many-body product state basis as in Eq.(\ref{monomial state}). The result is given by
\begin{equation}
\label{many-body matrix}
    \langle\phi_\lambda|\hat V|\phi_\mu\rangle=
    \begin{cases}
        \sum_{i=1}^NV^{1bdy}_{\lambda_i\lambda_i} &, \lambda=\mu\\
        (-1)^{a+b}V_{\lambda_a\mu_b} &,\lambda^{[a]}=\mu^{[b]}\\
        0 &,\text{ otherwise}
    \end{cases}
\end{equation}
where we use the notation $\lambda^{[a]}$ to denote the partition obtained by removing the $a$-th part in $\lambda$, i.e. $\lambda^{[a]}=(\lambda_1,\lambda_2,...,\lambda_{a-1},\lambda_{a+1},....,\lambda_N)$. The matrix element $\langle\phi_\lambda|\hat V^{1bdy}|\phi_\mu\rangle$ is only non-zero of $\lambda$ and $\mu$ are the same partition (i.e. it is a diagonal term) or if $\lambda$ and $\nu$ differ by a single element (since the potential is one-body).

\subsection{Berry phase calculation}
The ground state of any Hamiltonian can be found by exact diagonalization up to a random phase. The Berry phase can be obtained by calculating
\begin{equation}
\label{cumulative overlap}
\mathcal De^{i\gamma_{(N)}} = \prod_{i=1}^N\langle \psi(\theta_i)|\psi(\theta_{i+1}) \rangle, N+1\equiv 1
\end{equation}
where $|\psi(\theta_i)\rangle$ is the ground state of $H(\theta_i)$ and $0=\theta_1,\theta_2,....,\theta_N=2\pi$ are points along the closed loop in parameter space. The final term, $\langle\psi(\theta_{N})|\psi(\theta_1)\rangle$ is added to remove the random gauge that results from diagonalizing each $H(\theta_i)$ independently so that the quantity in Eq.(\ref{cumulative overlap}) is gauge-independent. In the limit $N\to\infty$, $\mathcal D\to1$ and $\gamma_{(N)}\to\gamma$, the total adiabatic phase accumulated. In practice, for a finite $N$, the closeness of $\mathcal D$ to unity indicates how well $\gamma_{(N)}$ approximates the Berry phase\cite{wang2019lattice}. All numerics presented here are done on the disk geometry.

\subsection{Berry curvature}
The Berry curvature at any point $\mathbf{r}$ can be calculated by evaluating the Berry phase around a small loop of radius $\epsilon$ around $\mathbf{r}$. The choice of radius $\epsilon$, as well as the number of evaluation point $N$ to apply Eq. (\ref{cumulative overlap}) depends on the Berry curvature gradient at the specific point. In general, a more irregular Berry curvature requires a smaller loop with more evaluation points. The amplitude $\mathcal D$ gives an indication whether the result of the calculation is reliable.

\section{Numerical results}
\subsection{Exchanging two elliptical holes}
For the IQH state on the LLL, each quasihole is a hole with fermionic statistics. For two squeezed IQH holes, one can verify that the fermionic statistics is observed for any squeezing ratio provided that the exchange procedure does not rotate them. However, when the squeezed holes are exchanged with self-rotation, we see deviation from the fermionic statistics (see Fig.\ref{fig:rotate ellipse}). Here the exchange statistics is taken to be the values of the plateau formed in the middle at separation distance between 6 and 8. A separation too small means the quasiholes are not well separated, while on a finite systems a separation too large puts the quasiholes too close to the edge of the quantum Hall droplet. In Fig.\ref{fig: ellipse density}, only figure b shows two quasiholes that are both separate from each other and far away from the edge of the disk. In numerics we are only interested in this region.

To tell the exchange statistics apart from such finite size effect, we plot the exchange phase (difference in the total pha se from the enclosed area) against separation distance, as in Fig.\ref{fig:rotate ellipse}b. On a large enough disk it is possible to observe a range of separation in which the exchange phase is relatively constant. In Fig.\ref{fig:rotate ellipse}, we take the exchange phase to be the values averaged within the range $R\in(6,8)$. This value decreases from 0.5 as the squeezing ratio is increased. As shown in the main text, the amount of deviation agrees well with the phase gained by a single elliptical hole rotating about its center by $2\pi$. This lead us to conclude that nontrivial self-rotation phase results in additional contribution to the quasihole statistics apart from the topological spin. Both contributions are captured by the proposed intrinsic spin.

\subsection{Toy model for realistic condition}
The fact that deformation of a quasihole can modify the braiding statistics can have consequences in real experiments. Here we illustrate how the connection between anyonic statistics and rotation can help us understand and improve experimental measurement. We consider a toy model consisting of two Diract delta potential traps, each pinning a quasihole, on a disk with a potential ``wall". 

\begin{figure}
\begin{center}
\includegraphics[width=\linewidth]{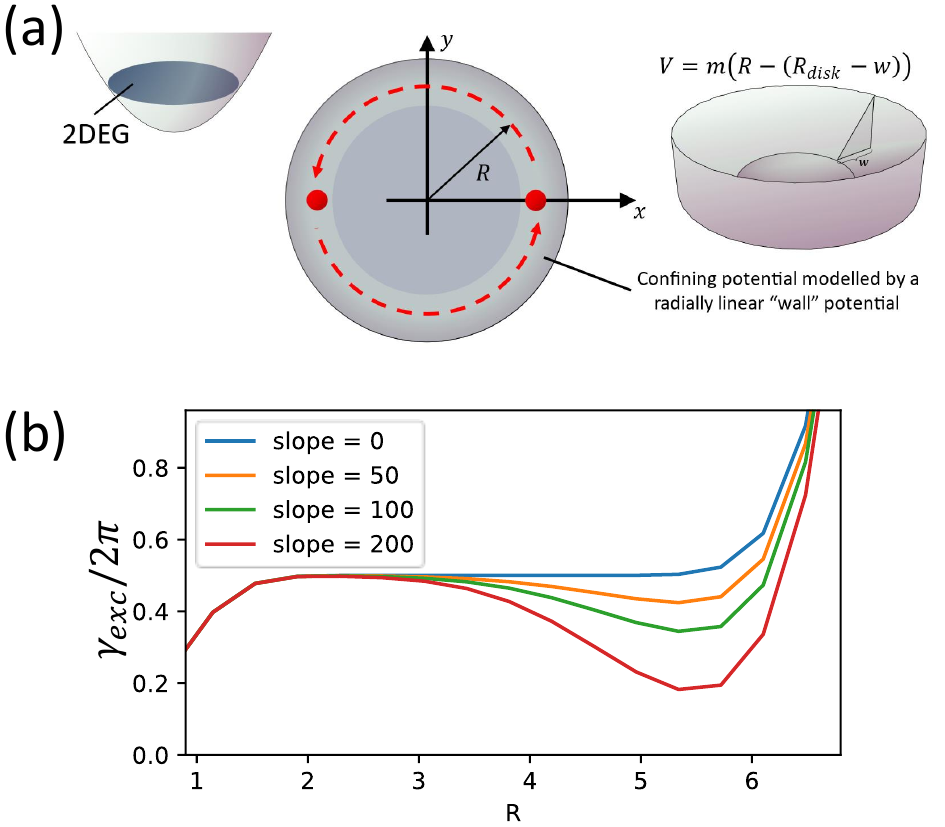}
\caption{(a) Modelling a 2DEG in a confining potential with a ``wall" potential on the quantum Hall droplet. The two IQH holes (red solid circles) are exchanged along a circle of radius $R$ (dashed arrows). (b) Statistical phase of two IQH holes when exchanged along a circle of radius $R$, for wall potential of width $W=3$ and varying slopes}
\label{fig:wall potential}
\end{center}
\end{figure}

We consider the following potential profile
\begin{equation}
\label{potential with trap}
V(z) = \delta^2(z-Re^{-i\theta}) + \delta^2(z-Re^{i\theta}) + V_{wall}(z)
\end{equation}
consisting of two Diract delta potential traps, each pinning a quasihole, on a disk with a potential ``wall". The wall potential takes the form
\begin{equation}
\label{wall potential}
V_{wall}(z) = m(|z|-(R_{disk}-W)
\end{equation}
which increases linearly from 0 between $|z| = R_{disk}-W$ and $R_{disk}$. Here $R_{disk}$ is the radius of the quantum Hall droplet, $m$ is the slope, and $W$ is the width. This potential approximates a potential trap used in real experiments to contain the entire quantum Hall droplet (see Fig.\ref{wall potential}a). The potential wall in our model increases linearly in the radial direction, from $V(|z|=R_{disk}-w)=0$ to $V(|z|=R_{disk})=mw$. (This potential wall is parametrized by the width $w$ and the slope $m$, see Fig. \ref{fig:wall potential}a.)

Each of the Dirac delta potential pins one quasiholes, and they can be easily manipulated along an exchange path of radius $R$. When $R\sim R_{disk}$, this setup emulates experiments where the braiding anyons are carried by the edge current\cite{nakamura2020direct, bartolomei2020fractional}. Fig.\ref{fig:wall potential}b shows the exchange phase at different values of $m$. The case when $m=0$ presents the ideal scenario where there is no extra potential on the QH droplet beside the pinning potentials. Here we see a relatively flat plateau between $R=2$ and $R=5$ at $\gamma_{exc}/2\pi=0.5$. This is the expected statistics of the hole at $\nu=1$. The region $R>5$ where the exchange phase increases presents the finite size effect where the quasihole is deformed by the edge of the Hall droplet.

When the slope of the wall potential is finite, there is an additional deformation to the quasihole which leads to changes in the statistics, seen between $R=4$ and $R=5$ (refer to the main text for the density plot of the quasihole here). We emphasize that this deviation is distinct from the finite size effect and hence must be accounted for when considering a realistic setup. In this toy model, the exchange phase at $m=0$ can be linearly extrapolated from the values at $m>0$, as discussed in the main text. How well this toy model corresponds to real experiment conditions remains to be studied, but in general we expect any potential used to trap the electron gas to have a measurable effect on the exchange statistics near the edge.

\section{Composite Fermionization between conformal Hilbert spaces}
\subsection{The CF formalism}
In this section we will discuss a powerful tool of dealing with FQH phases, namely the composite fermionization method (fermionization for short)\cite{yang2022composite}. In particular we will focus on the isomorphism between the Hilbert spaces before and after the fermionization, which allows us to define the angular momentum within a specific CHS as shown in the main text.

As a phenomenological theory, composite fermionization\cite{jain2007composite} combines one electron with even number of magnetic fluxes into a new bound state named as composite fermion (CF), which will significantly decrease the difficulty of resolving the interactions between particles and transform a strong-coupling FQH phase of electrons to some effective weak-interacting phase (such as an IQH phase) of CFs. Besides the vivid electron-flux binding illustrations, the CF picture can also provide the microscopic wave function for many states. The validity of the whole CF theory is built upon the assumption that the flux quanta are attached to electrons with a specific ratio and the gap will not close during the whole process of flux binding, neither of which is easy to be proven generically, though. However, inspired by the idea of fermionization, one can also focus on some aspects of the Hilbert space to pursue a more rigorous definition of fermionization\cite{yang2022composite}. Below we will show one approach from the invariant counting pattern of $L_z$ sectors.

Here we take the CF IQH state with $\nu^*=1$ and the Laughlin state of electrons with $\nu=1/3$ on the disk as an example. The generating function of the degeneracy of the $L_z$ sector with $N_e$ electrons and $N^e_o$ orbitals within $\mathcal{H}_1$ (nullspace of $\hat{V}_1$) is given by:
\begin{equation}
\mathcal{G}_e(q)= q^{\frac{3 N_e (N_e-1)}{2}}
\left[\begin{array}{c}
N^e_o-2N_e+2 \\
N_e
\end{array}\right]_{q}
\end{equation}
Here $q$ is just a dummy variable whose coefficient gives the counting of the sector with $L_z$ equal to its power. Meanwhile the generating function with $N_{CF}$ CFs and $N^{CF}_o$ orbitals in the lowest CF level can be written as:
\begin{equation}
\mathcal{G}_{CF}(q)= q^{\frac{N_{CF} (N_{CF}-1)}{2}}
\left[\begin{array}{c}
N^{CF}_o \\
N_{CF}
\end{array}\right]_{q}
\end{equation}
Thus to make the Hilbert spaces isomorphic with respect to the $L_z$ sectors, we should look for a specific relation between $\{ N_e, N^e_o\}$ and $\{ N_{CF}, N^{CF}_o\}$ that makes the following relation holds:
\begin{equation}
\mathcal{G}_e(q) = \mathcal{G}_{CF(q)} \cdot q^n, \quad n\in \mathbb{Z}
\end{equation}
which means the $L_z$ sector countings of electrons and CFs are identical up to a shift of $L_z$. An easy observation is that if we have:
\begin{equation}
N_e = N_{CF}; N^{CF}_o = N^{e}_o -2N_e +2
\label{relations}
\end{equation}
then the CHS of electrons and the corresponding CF Hilbert space will be isomorphic, i.e.
\begin{equation}
\mathcal{H}_1(N_e, N^e_o) \cong \mathcal{H}_{LLL}(N_{CF}, N^{CF}_o)
\end{equation}
Note that the second equation in Eq.\ref{relations} suggests that we should combine each electron with two fluxes, which is exactly the same as the statement from the original CF theory. Moreover two additional orbitals should be added for the CFs, which will make the isomorphism holds for systems of \textit{any} sizes rather than only the thermodynamic limit. This formalism can be used for other geometries as well, for example it can naturally show that the fermionization commutes with angular momentum operators on the sphere.

One of the simplest cases is the fermionization of the squeezed Laughlin state at $\nu=1/3$ with a single quasihole. The squeezing will mix different $L_z$ sectors so we can generically express this state as:
\begin{equation}
|\psi_e \rangle = \sum_{m, \alpha} c_{m,\alpha} |m, \alpha \rangle_e
\end{equation}
where $m$ is the $L_z$ quantum number and $\alpha$ denotes the degeneracy in each $L_z$ sector, which will be $1$ within $\mathcal{H}_1$. Thus there is a one-to-one correspondence between each $L_z$ sector of electrons' and CFs' Hilbert spaces and all we need to do is to replace $|m, 1 \rangle_e$ with the corresponding $|m, 1 \rangle_{CF}$. For the general case of $\alpha > 1$, the fermionization is given by a unitary transformation between the Hilbert spaces of electrons and CFs.

\subsection{Calculating the topological spin}
The topological spin of an FQH quasihole is defined in the main text as the component of the intrinsic angular momentum within its CHS. In practice it is difficult to calculate this quantity exactly, as compared to the cyclotron angular momentum $\hat L_{cy} = -\hat a^\dagger \hat a+1/2$ or the guiding center momentum $\hat L_{gc} = \hat b^\dagger b+1/2$, there is no expression for the CHS angular momentum in terms of the single-particle cyclotron operators $\hat a^\dagger$, $\hat a$ and guiding center operators $\hat b^\dagger$, $\hat b$. (This fact is a simple example of how within a CHS the fundamental degree of freedom is no longer single electrons, but involves many-electron interactions.) However, for Abelian FQH states, the fermionization process provides a simple way to numerically compute the CHS angular momentum by mapping it to the guiding center angular momentum of CFs.

Here we take the Laughlin quasihole as an example. The quasihole is ensured to reside completely within the Laughlin CHS by taking the limit $V_1^{2bdy}\to\infty$. As shown in Eq.(\ref{general Berry connection 3}), rotation of an arbitrary state $|\psi\rangle$ yields a phase described by a Berry connection that equals the difference in total angular momentum between that state and its rotationally invariant part $|\psi_0\rangle$. We can define a CHS angular momentum, $L_z^{V_1}$ that satisfies:
\begin{equation}
\label{V_1 am}
\langle \psi|L_z^{V_1}|\psi\rangle - \langle\psi_0|L_z^{V_1}|\psi_0\rangle = \langle\psi|L_z^{gc}|\psi\rangle-\langle\psi_0|L_z^{gc}|\psi_0\rangle
\end{equation}
We define such an operator with the help of the fermionization process described above:
\begin{equation}
\label{V_1 am definition}
\langle\psi|L_z^{V_1}|\psi\rangle\eqdef\langle\psi|\CF \bar L_z|\psi\rangle\CF , \forall |\psi\rangle\in \mathcal H_1
\end{equation}
where $|\psi\rangle_{CF}$ is the composite-fermionized counterpart of $|\psi\rangle$. It is important to note that the CF mapping does not mix different $L_z$ sectors, hence Eq.(\ref{V_1 am definition}) is well-defined.

In general, a state living in $\mathcal H_1$ can be decomposed into different $L_z$ sectors:
\begin{equation}
\label{state}
|\psi\rangle = \sum_m c_m|\psi_m\rangle
\end{equation}
where $L_z|\psi_m\rangle = \ell_m|\psi_m\rangle$. In first quantization, each $\psi_m$ may be a linear combination of different Jack polynomials within the same $L_z$ sector. Since the CF mapping does not mix different $L_z$ sectors, we can write the corresponding CF state in a similar form:
\begin{equation}
\label{state CF}
|\psi\rangle\CF  = \sum_m c_m|\psi_m\rangle\CF
\end{equation}
where we now have $L_z|\psi_m\rangle\CF  = \tilde{\ell}_m|\psi_m\rangle\CF $. $\tilde{\ell}_m$ and $\ell_m$ are related by a constant shift, so we have $\bar\ell_m-\bar\ell_0 = \ell_m-\ell_0$. Substituing these states into Eq. (\ref{V_1 am}) using the definition in Eq.(\ref{V_1 am definition}), and noting that $|\psi_0\rangle=|\psi_{m=0}\rangle$ and $\sum_m|c_m|^2 = 1$, we obtain:
\begin{align}
\langle\psi|\CF L_z|\psi\rangle\CF -&\langle\psi_0|\CF L_z|\psi_0\rangle\CF\nonumber\\
&=\sum_{m\geq 0}|c_m|^2\tilde\ell_m - \tilde\ell_0\label{proof1}\\
&=\sum_{m>0}|c_m|^2\left(\tilde\ell_m-\tilde\ell_0\right)\label{proof2}\\
&=\sum_{m>0}|c_m|^2\left(\ell_m-\ell_0\right)\label{proof3}\\
&=\sum_{m\geq0}|c_m|^2\ell_m - \ell_0\label{proof4}\\
&=\langle\psi|L_z|\psi\rangle-\langle\psi_0|L_z|\psi_0\rangle
\end{align}
Thus, we see that Eq.(\ref{V_1 am}) is satisfied by the definition in Eq.(\ref{V_1 am definition}).

Eq.(\ref{V_1 am definition}) can be understood as a result of treating the CF as the fundamental degree of freedom within the nullspace of $\hat V_1^{2bdy}$. Therefore, rotating a Laughlin quasihole within $\hat V_1^{2bdy}$ nullspace can be viewed as rotating a CF hole within a single CF level, which is exactly analogous to rotating a hole within the single LL. With this understanding, the discussions of the braiding properties with numerical evidences on the IQH states presented in the main text and in the sections above can be generalized to all Abelian states. In particular, metric deformation within a CHS leads to a change in the intrinsic spin, which in turns affect the exchange statistics according to our generalized spin-statistics relation.

\bibliographystyle{apsrev}
\bibliography{ref}
\end{document}